\documentclass[prd,aps,amssymb,amsmath,amsfonts,11pt]{revtex4}
\usepackage{latexsym}
\usepackage[latin1]{inputenc}
\usepackage[dvips]{graphicx}
\usepackage{color}

\usepackage{ulem}
\usepackage{epsfig}

\newcommand{\ba}{\begin{eqnarray}}
\newcommand{\ea}{\end{eqnarray}}

\newcommand{\be}{\begin{equation}}
\newcommand{\ee}{\end{equation}}

\newcommand{\HQSS}{{\rm HQSS}}
\newcommand{\SU}{\mbox{SU}}
\usepackage{dcolumn}

\begin{document}

\title{Propagation of heavy baryons in heavy-ion collisions} 
\author{Santosh K. Das$^{a,b}$, Juan M. Torres-Rincon$^{c,d}$, Laura Tolos$^{e,d}$, Vincenzo Minissale$^{a,b}$,
 Francesco Scardina$^{a,b}$, Vincenzo Greco$^{a,b}$}
\affiliation{$^a$ Department of Physics and Astronomy, University of Catania,
Via S. Sofia 64, 1-95125 Catania, Italy} 
\affiliation{$^b$ Laboratori Nazionali del Sud, INFN-LNS, Via S. Sofia 62, I-95123 Catania, Italy }
\affiliation{$^c$ Subatech, UMR 6457, IN2P3/CNRS, Universit\'e de Nantes, \'Ecole de Mines de Nantes, 4 rue Alfred Kastler 44307,
Nantes, France
\\
$^d$ Frankfurt Institute for Advanced Studies. Johann Wolfgang Goethe University, Ruth-Moufang-Str. 1,
60438, Frankfurt am Main, Germany}
\affiliation{$^e$ Institut de Ci\`encies de l'Espai (IEEC/CSIC), Campus Universitat Aut\`onoma
de Barcelona, Carrer de Can Magrans s/n, 08193 Cerdanyola del Vall\'es, Spain}

\begin{abstract}
The drag and diffusion coefficients of heavy baryons ($\Lambda_c$ and $\Lambda_b$) in the 
hadronic phase created in the latter stage of the heavy-ion collisions at RHIC and LHC energies have been evaluated recently.
In this work we compute some experimental observables, such as the nuclear suppression factor $R_{AA}$ and the elliptic
flow $v_2$  of heavy baryons at RHIC and LHC energies, highlighting the role of the hadronic phase contribution to these 
observables, which are going to be measured at Run 3 of LHC. For the time evolution of the heavy quarks in the QGP and heavy 
baryons in the hadronic phase we use the Langevin dynamics. For the hadronization of the heavy quarks to heavy baryons 
we employ Peterson fragmentation functions. We observe  a strong suppression of both 
the $\Lambda_c$ and $\Lambda_b$. We find that 
the hadronic medium has a sizable impact on the heavy-baryon elliptic flow whereas the impact of hadronic 
medium rescattering is almost unnoticeable on the nuclear suppression factor. 
We evaluate the $\Lambda_c/D$ ratio at RHIC and LHC. We find that $\Lambda_c/D$ ratio remain unaffected due to 
the hadronic phase rescattering which enable it as a nobel probe of QGP phase dynamics along with its hadronization.
\end{abstract}

\maketitle
\newpage
\tableofcontents

\section{Introduction}

 Ongoing experiments on relativistic heavy-ion collisions at high energies, like the Relativistic Heavy Ion Collider
(RHIC) and the Large Hadron Collider (LHC), have been designed to reach a new state of matter known as quark and gluon
plasma (QGP). It is a deconfined strongly-interacting plasma behaving like an almost perfect fluid. The bulk properties of this
high-temperature phase are governed by the light quarks and gluons. However, charm and bottom quarks (collectively denoted as heavy quarks) are
responsible for several observables which are essential to probe the QGP properties. The reason is that these heavy quarks are witnesses of
the entire plasma evolution as they are produced in the initial hard scatterings and remain abiding until hadronization.
In their final state they appear as constituents of heavy hadrons, mainly $D$ and $B$ mesons. Indeed, these states have generated significant interest in the
recent past because they serve as indicators of QGP dynamics~\cite{Prino:2016cni} due to the suppression of their momentum distribution at large $p_T$ in the thermal medium, reflected
in a low nuclear suppression factor $R_{AA}$, and a sizable value of the elliptic flow $v_2$, a measure of the azimuthal 
anisotropy in the plasma. Noticeably $R_{AA}$ and $v_2$ have similar values like the light hadrons.

Thanks to the last upgrades in the experimental detectors, RHIC and LHC can reconstruct $D$ mesons from their hadronic decay products (like $D^0 \rightarrow K^- \pi^+$),
instead of collecting nonphotonic electrons coming from semileptonic decays. With recent experimental results from STAR (RHIC)~\cite{Adamczyk:2014uip} as well
as ALICE (LHC)~\cite{ALICE:2012ab,Abelev:2013lca,Abelev:2014ipa,Adam:2015sza}, one can now contrast the predictions of the theory groups, which have computed 
the $R_{AA}$ and $v_2$ of heavy mesons, using numerical simulations for the heavy-ion evolution under different models~\cite{Linnyk:2008hp,Das:2010tj,
Mazumder:2011nj,Lang:2012yf,He:2012df,He:2011qa,Das:2013kea,Alberico:2011zy,Uphoff:2011ad,Uphoff:2012gb,Cao:2013ita,Cao:2015hia,
Gossiaux:2008jv,Nahrgang:2013xaa,Nahrgang:2014vza,Ozvenchuk:2014rpa,Song:2015sfa,Song:2015ykw}. 

 In addition to heavy mesons, future upgrades in the ALICE detector will allow to study $\Lambda_c$ and $\Lambda_b$ baryons within the so-called Run 3 of
LHC~\cite{Abelevetal:2014dna,Andronic:2015wma,Dainese:2016dea} (see~\cite{Aaij:2013mga} for a recent study on the $\Lambda_c$ baryon 
reconstruction in $p+p$ collisions by the LHCb
collaboration). As presented in Ref.~\cite{Abelevetal:2014dna}, the ALICE collaboration plans to study several observables related to $\Lambda_c$ baryons, namely
the $R_{AA}$, $v_2$ and $\Lambda_c/D$ ratio. Given some key upgrades in the ALICE detector capabilities, the $\Lambda_b$ physics in heavy-ion collisions has
also been considered for the Run 3~\cite{Abelevetal:2014dna,Andronic:2015wma}.

These experimental advances on heavy baryons ($\Lambda_c$ and $\Lambda_b$) are of general interest because they will allow us to
have novel information on the hadronization mechanism and, more specifically, on the evaluation of the heavy
baryon-to-meson ratio ~\cite{Oh:2009zj,Lee:2007wr}. In the light and strange sectors, this ratio has shown an anomalous enhancement
with respect to $p+p$ collisions.
Also an enhancement would affect the $R_{AA}$ of non-photonic electrons~\cite{Sorensen:2005sm,Ayala:2009pe}. This is because the branching 
ratio of heavy-baryon decay~\cite{MartinezGarcia:2007hf} to electrons is smaller than the branching ratio of heavy meson 
to electrons. Furthermore the heavy baryon to meson ratio, ($\Lambda_c/D$ and $\Lambda_b/B$), is very 
fundamental for the understanding of the in-medium hadronization~\cite{Greco:2003vf,Fries:2008hs}.

 In a recent work~\cite{Tolos:2016slr} some of us have extensively studied the microscopical details of the $\Lambda_c$ and
$\Lambda_b$ interactions with light mesons, such as $\pi,K,\bar{K},\eta$ (see also Ref.~\cite{Ghosh:2014oia} for a
first study in this direction). In Ref.~\cite{Tolos:2016slr} the authors used an effective field theory at low energies to
describe the hadronic interactions, which are, in addition, unitarized to account for the required unitarity property of the 
scattering amplitudes. They presented the typical cross sections for both $\Lambda_c$ and $\Lambda_b$ baryons, containing 
many resonant states. Then, the authors computed the relevant transport coefficients, drag force and diffusion coefficients, as a function of the
temperature and heavy baryon momentum for the conditions expected after the hadronization in high-energy  heavy ion collisions. 
Either the individual cross sections, or the
transport coefficients themselves, can be readily used in transport simulations to account for the $R_{AA}$ or $v_2$ of heavy 
baryons at heavy-ion collisions at RHIC and LHC.

 As an application of the findings in~\cite{Tolos:2016slr}, we now present predictions for
$R_{AA}$ as well as $v_2$  of $\Lambda_c$ and $\Lambda_b$ baryons for RHIC and LHC energies for  an eventual
comparison  with experimental results and to understand
if we can describe both the heavy meson and heavy baryon observables
simultaneously. We also present prediction for the 
$\Lambda_c/D$ ratio for RHIC and LHC energies. We will accommodate a Langevin equation 
for the momentum evolution of the heavy particles (equivalent to a Fokker-Planck realization) whose parameters are related to the 
drag and diffusion coefficients. These are taken from a quasiparticle model~\cite{Das:2015ana} for the heavy quark 
propagation, and from Ref.~\cite{Tolos:2016slr} for the hadronic phase.

This manuscript is organized as follows. In Sec.~\ref{sec:langevin} we describe the Langevin equations to be solved for the dynamics of the heavy particle.
In particular, we explain how the coefficients of the equations of motion are related to the transport coefficients computed in our previous work~\cite{Tolos:2016slr}.
In Sec.~\ref{sec:model} we provide some details about the practical implementation of our model: we describe our prescription for the initial state, the quasiparticle
model used for the propagation of heavy quarks in the hot plasma, the hadronization mechanism for the confined phase transition, and the freeze-out condition.
Our results are presented in Sec.~\ref{sec:observables}, where we give our predictions for $R_{AA}$~(\ref{sec:raa}) and $v_2$~(\ref{sec:v2}). 
Sec.~\ref{sec:ratio} is devoted for heavy baryon to meson ratio. Finally, we draft our conclusions in Sec.~\ref{sec:conclusions}.

\section{Langevin equation for heavy particles~\label{sec:langevin}}

The standard approach to heavy-quark dynamics in the QGP and the propagation of open-heavy hadrons in the hadronic medium is to 
follow their evolution by means of a Fokker-Planck equation solved stochastically by the Langevin equations~\cite{Prino:2016cni}. 
The relativistic Langevin
equations of motion for the time evolution of the position and momentum of the 
heavy quarks/heavy hadrons can be written in the form
\be \left\{ \begin{array}{rcl}
 dx_i & = & \frac{p_i}{E}dt \ ,  \\
 dp_i & = & -F(p) p_i dt+C_{ij}(p)\rho_j\sqrt{dt} \ ,
 \label{lv1} \end{array}
 \right. 
\ee
where $dx_i$ and $dp_i$ are the shift of the coordinate and momentum in each discrete time step $dt$.
$F(p)$ and $C_{ij}(p)$ are the drag force and the
covariance matrix respectively. $\rho$ is the noise which obeys the probability distribution of independent Gaussian-normal 
distributed random variables, $P(\rho)=(2\pi)^{-3/2}e^{-\rho^2/2}$, along with 
the relations $<\rho_i \rho_j>=\delta_{ij}$ and $<\rho_i>=0$. 
The covariance matrix is related to the transverse and longitudinal diffusion coefficients, 
\begin{eqnarray}
C_{ij}=\sqrt{2\Gamma_0(p)} \Delta_{ij}+\sqrt{2\Gamma_1(p)} \ \frac{p_i p_j}{p^2} \ , 
\label{cmmm}
\end{eqnarray}
where $\Delta_{ij} = \delta_{ij}-p_i p_j/p^2$ is the transverse projector operator.
Under the assumption, $\Gamma_0 (p)=\Gamma_1 (p)=\Gamma (p)$, 
Eq~(\ref{cmmm}) reduces to $C_{ij}=\sqrt{2 \Gamma(p)} \delta_{ij}$. Such an assumption, strictly valid 
for $p\rightarrow 0$, is usually employed at finite $p$  for heavy quark dynamics 
in the QGP~\cite{Moore:2004tg,vanHees:2005wb,Cao:2011et,He:2012df,Das:2010tj,Mazumder:2011nj,Lang:2012yf}.
With the knowledge of $F(p)$ and $\Gamma(p)$ as functions of $T$ and $p$, the Langevin equation is ready to be solved.
We use pre-Ito discretization scheme for the numerical implementation of the Langevin dynamics.

\section{Dynamical model~\label{sec:model}}

To solve the Langevin equation in the QGP/hadronic phase one needs the drag and diffusion coefficients of 
heavy quarks/heavy baryons as a function of temperature and momentum
in the QGP/hadronic medium. The drag and diffusion coefficients of the heavy quarks in the QGP are calculated 
inspired by the quasi-particle model
(QPM)~\cite{Das:2012ck,Berrehrah:2013mua,Berrehrah:2014kba}. The quasi-particle approach accounts for 
the non-perturbative dynamics
by means of temperature-dependent quasi-particle masses for light quarks and gluons, respectively,
\ba m^2_q & = & \frac{2N_c + N_f}{12} g^2(T) T^2 \ , \\ 
 m^2_g & = & \frac{N_c^2-1}{8N_c} g^2(T) T^2 \ , \ea
as well as a $T$-dependent background field known as bag constant. 
The strong coupling constant is obtained by a fit of the lattice energy density and is parametrized as follows:
\be g^2(T)= \frac{48 \pi^2}{(11N_c-2N_f) \ln \left[ \lambda \left(T/T_c-T_s/T_c \right) \right]^2}\ , \ee
with $N_c=N_f=3$, $\lambda=2.6$ and $T_s/T_c=0.57$~\cite{Plumari:2011mk}.
The quasi-particle scheme is able to successfully reproduce the thermodynamics of lattice-QCD~\cite{Plumari:2011mk} 
by fitting the strong coupling $g(T)$. For the evaluation of the drag and diffusion coefficients in the QGP medium, we 
use the QPM approach recently addressed in Ref.~\cite{Das:2015ana} to describe heavy quark $R_{AA}$ and $v_2$ 
at RHIC and LHC energies. A self-consistent dynamical treatment should include the finite width of the quasiparticle, however 
the drag and diffusion are not significantly affected, see Ref~\cite{Berrehrah:2013mua,Berrehrah:2014kba}. 
The drag has been calculated in Ref.~\cite{Das:2015ana} and show a very mild $T$ dependence in comparison with 
perturbative QCD (pQCD) or AdS/CFT. We notice that a similar dependence is found in the T-matrix approach~\cite{vanHees:2007me,Riek:2010fk}.

\begin{figure}[ht]
\begin{center}
\includegraphics[width=17pc,clip=true]{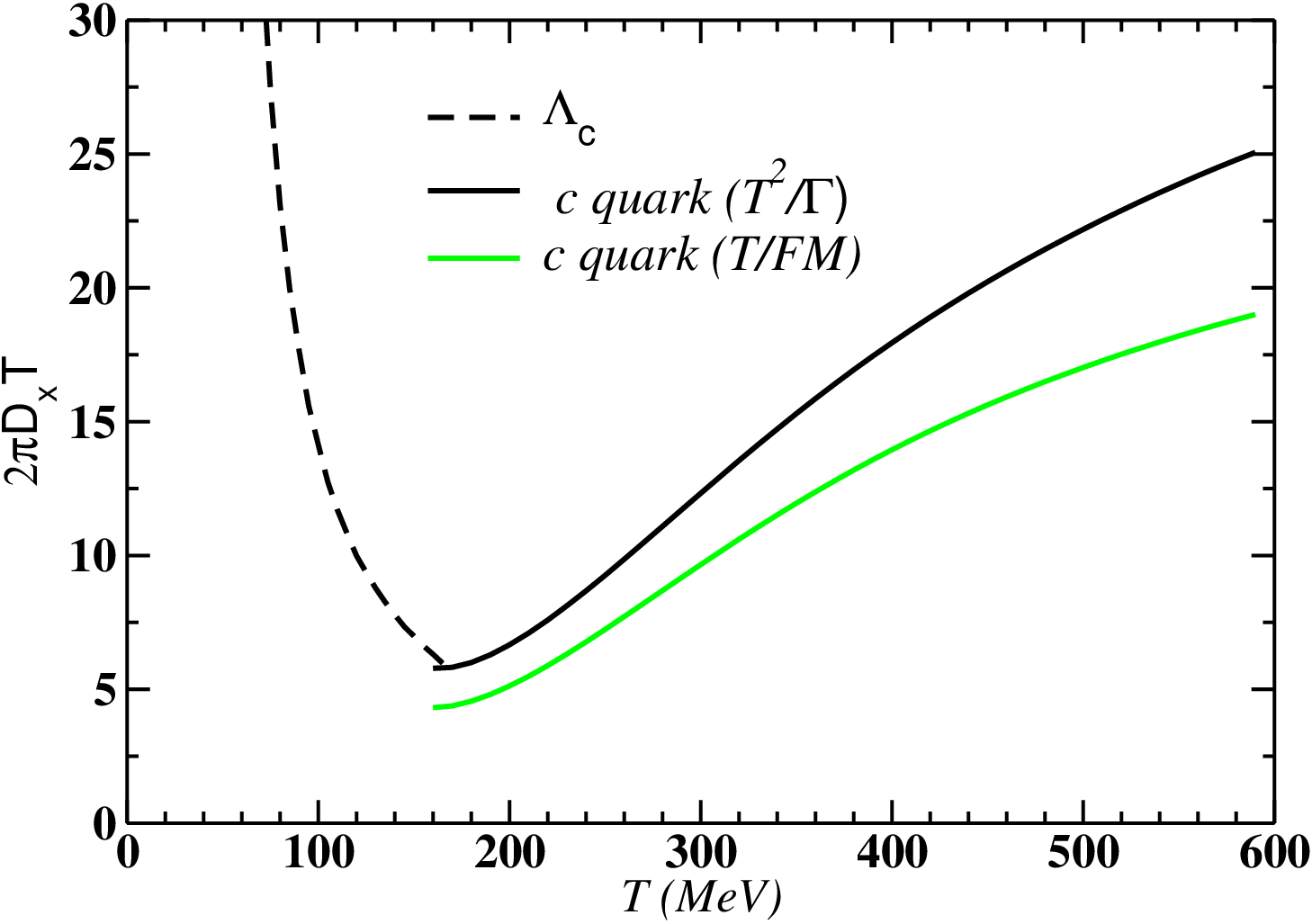}\hspace{2pc}
\includegraphics[width=17pc,clip=true]{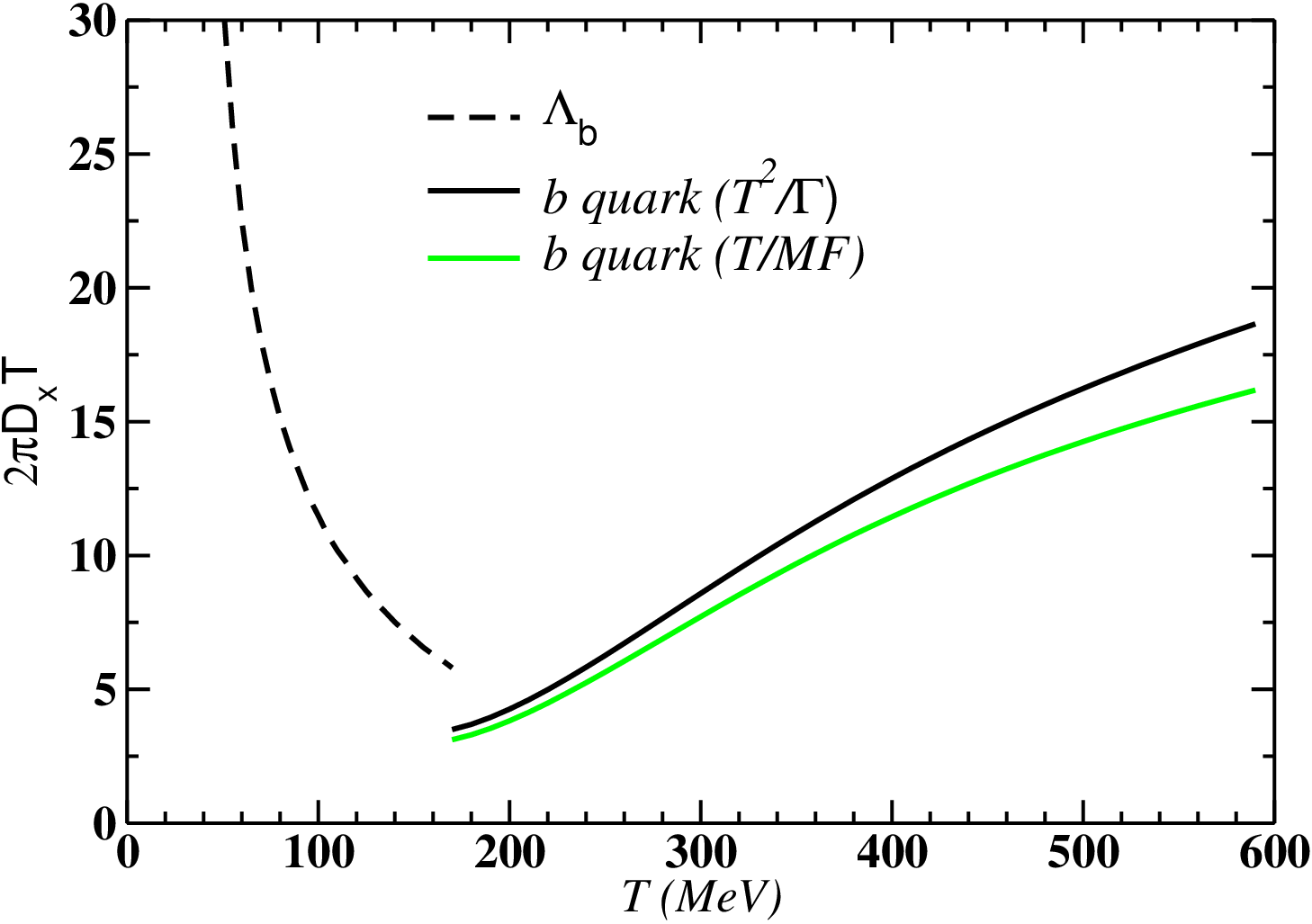}\hspace{2pc}
\caption{$D_x$ as a function of $T$ for $c$ quark and $\Lambda_c$ (left panel) and 
for $b$ quark and $\Lambda_b$  (right panel).}
\label{figx}
\end{center}
\end{figure}

Within the Fokker-Planck approach, the spatial  diffusion coefficient~\cite{Moore:2004tg,Abreu:2011ic}, $D_x$, 
can be calculated in the static limit ($p\rightarrow 0$) in two different ways. It can be obtained from the diffusion
coefficient in momentum space,
\be \label{eq:DxG} D_x = \frac{T^2}{\Gamma} \ , \ee
or from the drag coefficient using the Einstein relation ($\Gamma=MFT$),
\be \label{eq:DxF} D_x = \frac{T}{MF}  \ . \ee
However, the Einstein relation may not be strictly valid at high temperatures in the QGP phase. 
Hence, we are using both approaches (\ref{eq:DxG}) and (\ref{eq:DxF}) to evaluate the spatial diffusion coefficient in the QGP phase.

In Fig.~\ref{figx} the spatial diffusion coefficient in the QGP phase~\cite{Das:2015ana} is compared with the one for heavy baryons in hadronic matter~\cite{Tolos:2016slr}. 
In the left panel of this figure, we show $2\pi D_x T$ as a function of the temperature for $c$ quarks (high temperature) and $\Lambda_c$ baryons (low temperature).
We find that the $\Lambda_c$ diffusion coefficient also supports a continuous evolution with a minimum around $T_c$ like the heavy-meson case ($D$ meson)~\cite{He:2012df,Ozvenchuk:2014rpa,Berrehrah:2014tva}.
The differences of the $D_x$ in the QGP phase by the two different approaches are due to the violation of the Einstein relation. In the hadronic sector we have observed
that the Einstein relation is satisfied for all temperatures~\cite{Tolos:2016slr}.

In the right panel of Fig.~\ref{figx} we show $2\pi D_x T$ as a function of temperature for $b$ quarks (high temperature) and for $\Lambda_b$ baryons (low temperature).
In this case we also find an almost continuous evolution with a minimum around $T_c$. For the bottom case, the calculations of the spatial diffusion coefficient
using Eq.~(\ref{eq:DxG}) and Eq.~(\ref{eq:DxF}) are in better agreement than for the charm case. The reason is that due to the heavier mass of the $b$ quark, the Einstein relation is 
better satisfied than the charm case (violations are more severe at high temperature), and the two ways of computing $D_x$ are practically equivalent.

Once the temperature of the QGP phase goes below $T_c$, the QGP phase give way to the hadronic phase. 
In this phase heavy hadrons, produced after hadronization, suffer from collisions with light mesons. 
To fully characterize the QGP phase, the impact of hadronic phase should be then taken into account. Several attempts 
have been made in this direction to study the hadronic medium interaction  and their impact on heavy mesons ($D$ and $B$) 
observables at RHIC and LHC energies. However, little efforts have been given to the study of heavy 
baryon interaction in the hadronic phase. 

 Here, we use the recent results in Ref.~\cite{Tolos:2016slr}, where some of us analyzed the heavy baryon 
 interaction with the hadronic medium consisting of light mesons ($\pi$, $K$, $\bar K$ and $\eta$) within unitarized 
interactions from effective field theories that respect chiral and heavy-quark symmetries.
With these interactions, we have obtained the heavy-baryon transport coefficients (drag and diffusion) as a function 
of temperature and momentum. In the present work we aim at studying the heavy baryon evolution in the hadronic 
phase within the Langevin dynamics using the drag and diffusion coefficients calculated in the previous paper and highlight its 
impact on several observables potentially measurable at RHIC and LHC energies, in particular on $R_{AA}$ and $v_2$.

\subsection{Initialization and heavy quark dynamics}
The solution of the Langevin equation needs a background medium describing the space-time evolution of the bulk matter. 
To describe the expansion and cooling of the bulk matter and its elliptic flow $v_2(p_T)$ at both RHIC and LHC colliding 
energies, we have employed a 3D+1 relativistic transport code 
with an initial condition given by a standard Glauber model. Such a model allow us to describe the evolution of 
a fluid with a fixed $\eta/s$ in the same way as it is done by viscous hydrodynamical simulation. For more details we refer the 
reader to Refs.~\cite{Ruggieri:2013bda,Ruggieri:2013ova,Ferini:2008he,Greco:2008fs}.

In this work we have performed simulations of $Au+Au$ collisions at $\sqrt{s}= 200$ AGeV for the minimum bias. The initial
conditions for the bulk evolution in the coordinate space are given by the Glauber model. 
In momentum space we use a Boltzmann-J\"uttner distribution 
function up to a transverse momentum $p_T=2$ GeV, while at larger momenta mini-jet distributions as calculated within 
pQCD at Next-to-leading order (NLO) order~\cite{Greco:2003xt,Greco:2003mm}.

At RHIC energies for $Au+Au$ at $\sqrt{s}=200$ AGeV, the maximum initial temperature of the 
fireball at the center is $T_0=340$ MeV and
the initial time for the fireball simulations is chosen as $\tau_0=0.6$ fm/c (according to 
the criterion $\tau_0 \cdot T_0 \sim 1$ and to standard setting in hydrodynamics).  
We have also extended our calculation to study the heavy baryons $R_{AA}$ and $v_2$ at LHC energies 
performing simulations of $Pb+Pb$ collisions
at $\sqrt{s}= 5.5$ ATeV energy. In this case the maximum initial temperature at the center of 
the fireball is $T_0=610$ MeV  and the initial time for the
simulations is chosen as $\tau_0\sim 1/T_0 =0.25$ fm/c. We have performed simulations for $0-20\%$ centrality class. 

The heavy-quark distribution in momentum space, both for RHIC and LHC, is taken in accordance with the charm distribution 
in $p+p$ collisions, calculated within Fixed Order + Next-to-Leading Log (FONLL), taken from Ref.~\cite{Cacciari:2005rk,Cacciari:2012ny}, 
where in the coordinate space they are distributed according to number of binary nucleon-nucleon collisions ($N_{coll}$) 
from the Glauber model for both RHIC and LHC energies. We solve the Langevin dynamics to study the time evolution 
of heavy quark momentum in QGP created in $Au+Au$ collision as discussed in Sec.~\ref{sec:langevin}. The interaction 
between the heavy quarks and the bulk has been embedded through the drag and diffusion coefficients calculation within 
the QPM approach discussed at the beginning of this section.

\subsection{Hadronization and hadronic evolution\label{sec:frag}}
Another important aspect of a heavy-ion collision is the hadronization mechanism, when heavy quarks combine 
into color-neutral objects. Hadrons are formed when the temperature reaches $T=T_c=160$ MeV~\cite{Bazavov:2011nk}. 
One of the basic mechanisms of hadronization, widely 
considered in this context, is the fragmentation of an individual quark where the hadron momentum is a fraction $z$ 
of the quark momentum. For gluons and light quarks the fragmentation
functions are rather broad distributions around $z=0.5$, but for heavy quarks the fragmentation
functions  become rather sharply peaked towards $z=1$. 
The charm quark fragmentation for  $D$ meson and $\Lambda_c$ can be described using the 
Peterson fragmentation function~\cite{Pet},
\be
f(z) \propto 
\frac{1}{ z \lbrack 1- \frac{1}{z}- \frac{\epsilon_c}{1-z} \rbrack^2 } \ ,
\ee
where $\epsilon_c$ is a free parameter to fix the shape of the fragmentation function in comparison 
with the experimental data in $p+p$ collision. Unlike $D$ meson, the heavy baryon fragmentation function is not precisely 
known as it is yet to be measured in $p+p$ collisions. The $D$ meson spectra in $p+p$ collision at RHIC energy using FONLL 
calculation for 
the initial charm production can be reproduced using $\epsilon_c=0.01$. The $D$ meson spectra at LHC 
energy can be also reproduced using $\epsilon_c=0.01$. In the absence of the $p+p$ data for the $\Lambda_c$ production 
at RHIC and LHC energies, we are using the electron-positron annihilation data to fix the shape of the $\Lambda_c$ 
fragmentation. In electron-positron annihilation, the $\epsilon_c$ for the $\Lambda_c$ is  about a factor two larger
 than the $D$ meson one~\cite{hffs}. 
This means that the $\Lambda_c$ fragmentation function is softer than the $D$ meson fragmentation. 
This is because $\Lambda_c$ contains one heavy quark and two light quarks, whereas  
$D$ meson has one heavy quark and one light anti-quark. So in accordance with the electron-positron annihilation data,
we are using $\epsilon_c=0.02$ for the $\Lambda_c$, a factor two larger than the $D$ meson.

\begin{figure}[ht]
\begin{center}
\includegraphics[width=17pc,clip=true]{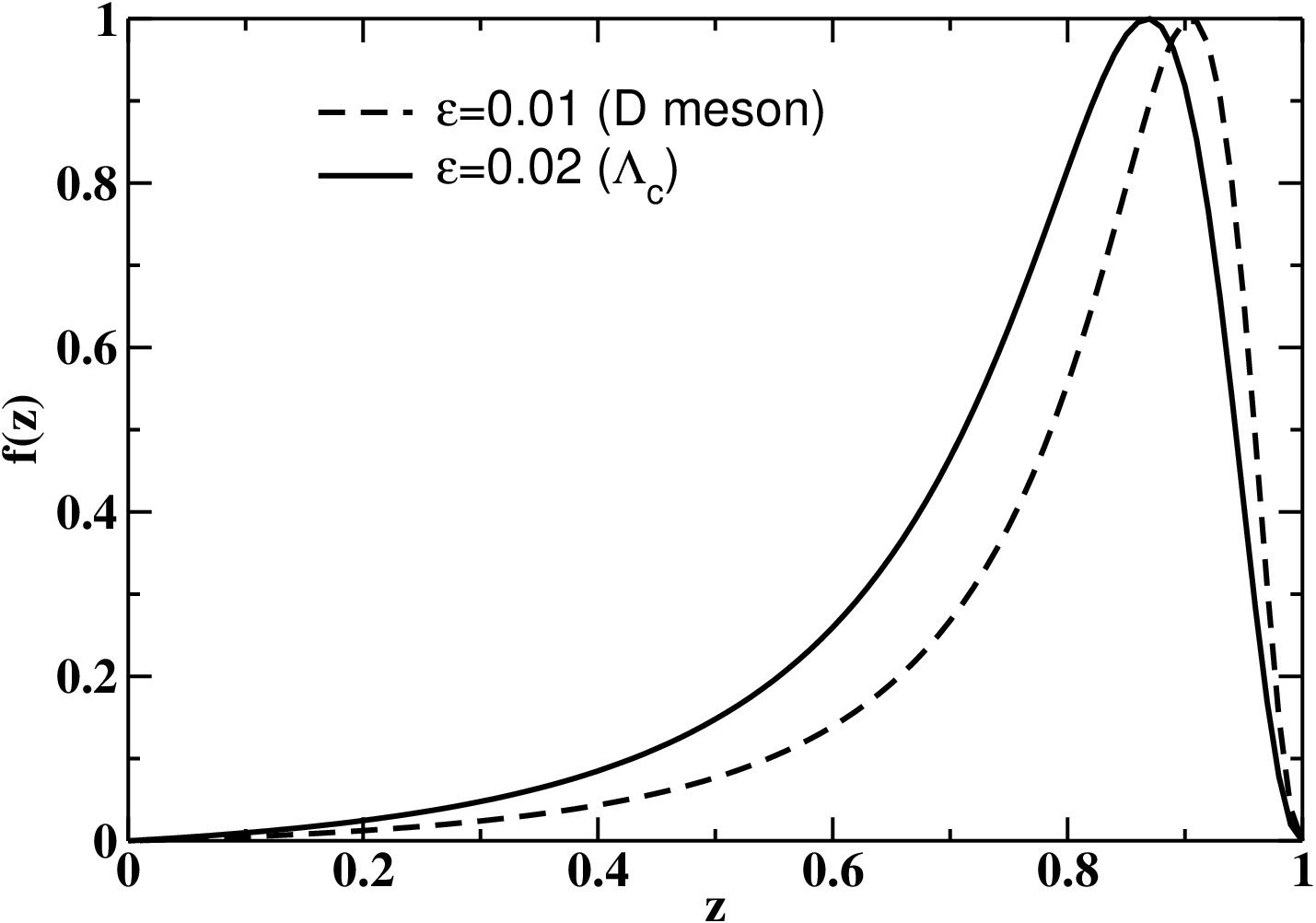}\hspace{2pc}
\caption{Variation of the fragmentation function with the fraction of momentum.}
\label{fig1}
\end{center}
\end{figure}

In Fig.~\ref{fig1} we show the variation of $\Lambda_c$ fragmentation function with the fraction of momentum  together with the 
$D$ meson fragmentation function. As expected, the  $\Lambda_c$ fragmentation function is softer than the $D$ meson as it
takes more energy to pop-up two quarks from the vacuum in the fragmentation picture.
For $\Lambda_b$ we use $\epsilon_c=0.006$, a factor two larger than the $B$-meson fragmentation function.

After the hadronization from the charm and bottom quarks to $\Lambda_c$ and $\Lambda_b$, respectively, 
we solve the Langevin dynamics for the propagation of $\Lambda_c$ and $\Lambda_b$ in an hadronic bath that consists 
of $\pi$, $K$, $\bar K$ and $\eta$. The interaction between the heavy baryons with the bath has been treated
within unitarized interactions based on effective field theories that respect chiral and heavy-quark symmetries. 
Specifically, the interaction of $\Lambda_c$ and $\Lambda_b$ scattering off $\pi$, $K$, $\bar K$ and $\eta$ mesons is obtained
 within a unitarized meson-baryon coupled-channel model that incorporates heavy-quark spin symmetry
~\cite{GarciaRecio:2008dp, Gamermann:2010zz, Romanets:2012hm,GarciaRecio:2012db,Garcia-Recio:2013gaa,Tolos:2013gta}.
This is a predictive model for four flavors including all basic hadrons (pseudoscalar and vector mesons, and $1/2^+$
 and $3/2^+$ baryons) which reduces to the Weinberg-Tomozawa interaction in the sector where Goldstone bosons
 are involved. This scheme has $\SU(6)\times \HQSS$ symmetry, i.e., spin-flavour symmetry in the 
 light sector and HQSS in the heavy (charm/bottom) sector, and 
 it is consistent with chiral symmetry in the light sector. For more details of the hadronic interaction 
we refer to the earlier work \cite{Tolos:2016slr}. The time evolution of heavy 
baryons within the hadronic phase is continued until the temperature reaches $T_{kin}=120$ MeV~\cite{Das:2013lra}, at 
the kinetic freeze out.

\begin{figure}[ht]
\begin{center}
\includegraphics[width=17pc,clip=true]{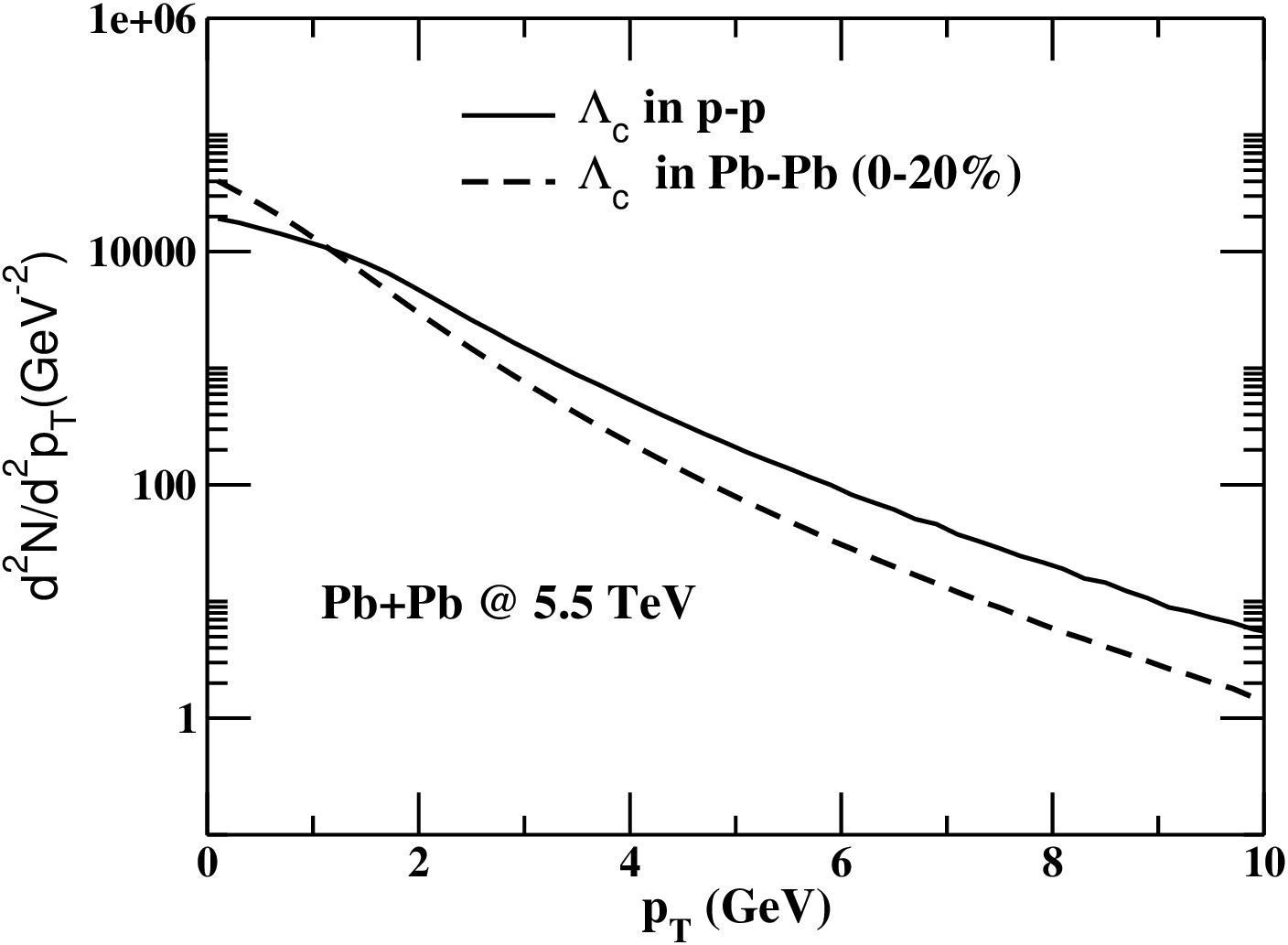}\hspace{2pc}
\caption{Variation of $\Lambda_c$ spectra in $p+p$ and $Pb+Pb$ collision at LHC colliding energy in arbitary normalization.}
\label{fig1_1}
\end{center}
\end{figure}

\begin{figure}[ht]
\begin{center}
\includegraphics[width=17pc,clip=true]{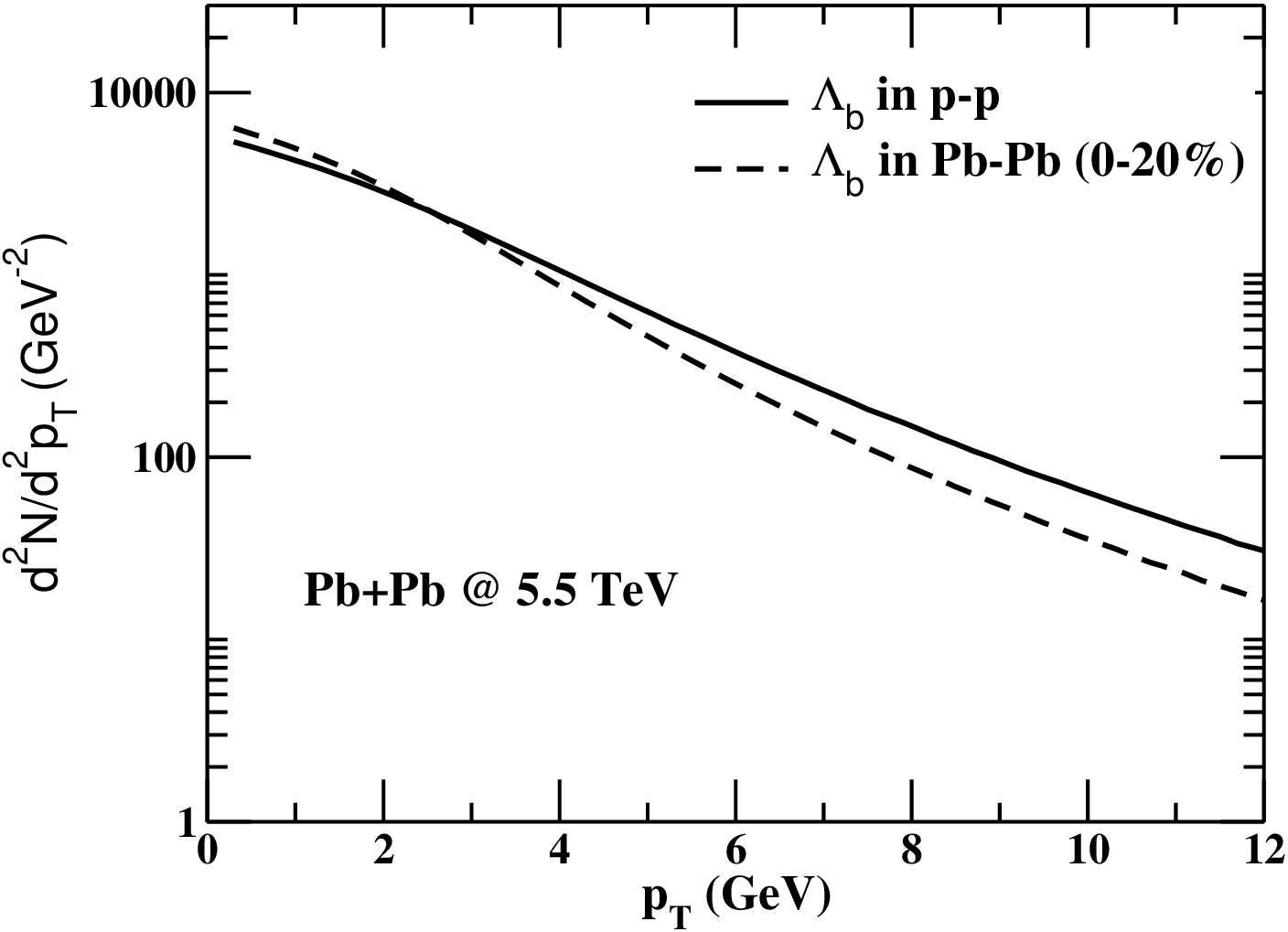}\hspace{2pc}
\caption{Variation of $\Lambda_b$ spectra in $p+p$ and $Pb+Pb$ collision at LHC colliding energy in arbitary normalization.}
\label{fig1_2}
\end{center}
\end{figure}

In Fig.~\ref{fig1_1} and Fig.~\ref{fig1_2} we show the variation of 
$\Lambda_c$ and $\Lambda_b$ spectra in $p+p$ and $Pb+Pb$ at LHC colliding energies in arbitary normalization.  
In the $Pb+Pb$ collision, due to interaction between heavy quarks and the bulk in the QGP phase as well as heavy baryons 
and the bulk in the hadronic phase, the heavy baryons rearrange their spectra with larger population at low momentum.  

\section{Results: experimental observables \label{sec:observables}}

The  heavy-baryon observables which are going to be measured at LHC Run 2 and 3~\cite{Abelevetal:2014dna,Tieulent:2015wec} 
are the nuclear suppression factor ($R_{AA}$) and the elliptic flow ($v_2$). We evaluate these
observables at both RHIC and LHC energies using Peterson fragmentation function as described above. One of our main motivations 
is to highlight the impact of the hadronic medium rescattering on heavy baryon observables.

\subsection{Nuclear modification factor, $R_{AA}$~\label{sec:raa}}

 One of the key observables related to heavy quark propagation, which is measured at RHIC and LHC energies,  
is the nuclear suppression factor $R_{AA}$. It measures the depletion of 
high transverse momentum ($p_T$) hadrons ($D$ and $B$ mesons) produced in nucleus+nucleus collisions with 
respect to those produced in 
proton+proton collisions scaled with the number of binary collision. 

The ALICE physics programme for Runs 3 and 4~\cite{Abelevetal:2014dna,Tieulent:2015wec} is going to measure the 
nuclear suppression factor $R_{AA}$ for heavy baryons. 
Keeping this in mind, we are keen to study the $R_{AA}$ of heavy baryons highlighting the possible impact of 
the hadronic medium. To access the effects of the QGP phase without hadronic interaction, 
we take the initial distribution of heavy quarks $f_i$ at $t=\tau_i$, and compare it with the distribution of 
heavy baryons right after the heavy quark fragmentation takes place ($f_{QGP\rightarrow HP}$ at $T_c$), that is
\be R_{AA}^{QGP \rightarrow HP}(p)=\frac{f_{QGP \rightarrow HP} (p)}{f_i (p)} \ . \ee
Similarly, the suppression factor in the hadronic phase alone can be written as 
\be R_{AA}^{HP}(p)=\frac{f_{HP}(p)}{f_{QGP \rightarrow HP }(p)} \ , \ee
where  $f_{HP}$ is the solution of the Langevin equation describing the evolution in the hadronic phase at 
the freeze out $T_{kin}=120$ MeV.
Notice that in the absence of any hadronic rescattering effect $R_{AA}^{HP}=1$.

The net suppression of the heavy mesons during the entire evolution process, from the beginning 
of the QGP phase to the end of the hadronic phase is given by:

\be
R_{AA}(p)=R_{AA}^{QGP \rightarrow HP}(p) \times R_{AA}^{HP}(p)= \frac{f_{HP} (p)}{f_i (p)} \ ,
\ee
which in the absence of genuine hadronic effects $R_{AA} (p) \simeq R_{AA}^{QGP \rightarrow HP} (p)$.

\begin{figure}[ht]
\begin{center}
\includegraphics[width=17pc,clip=true]{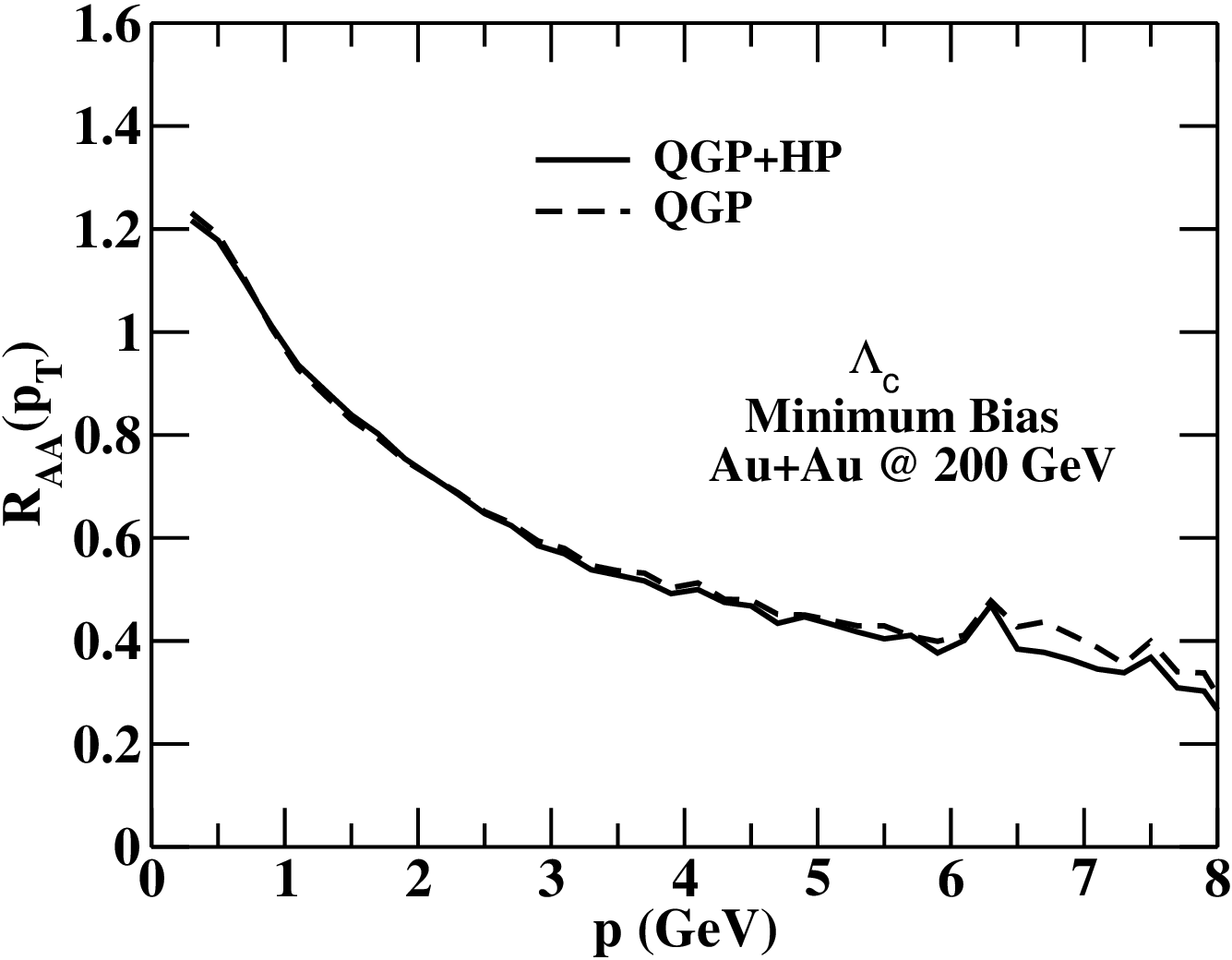}\hspace{2pc}
\includegraphics[width=17pc,clip=true]{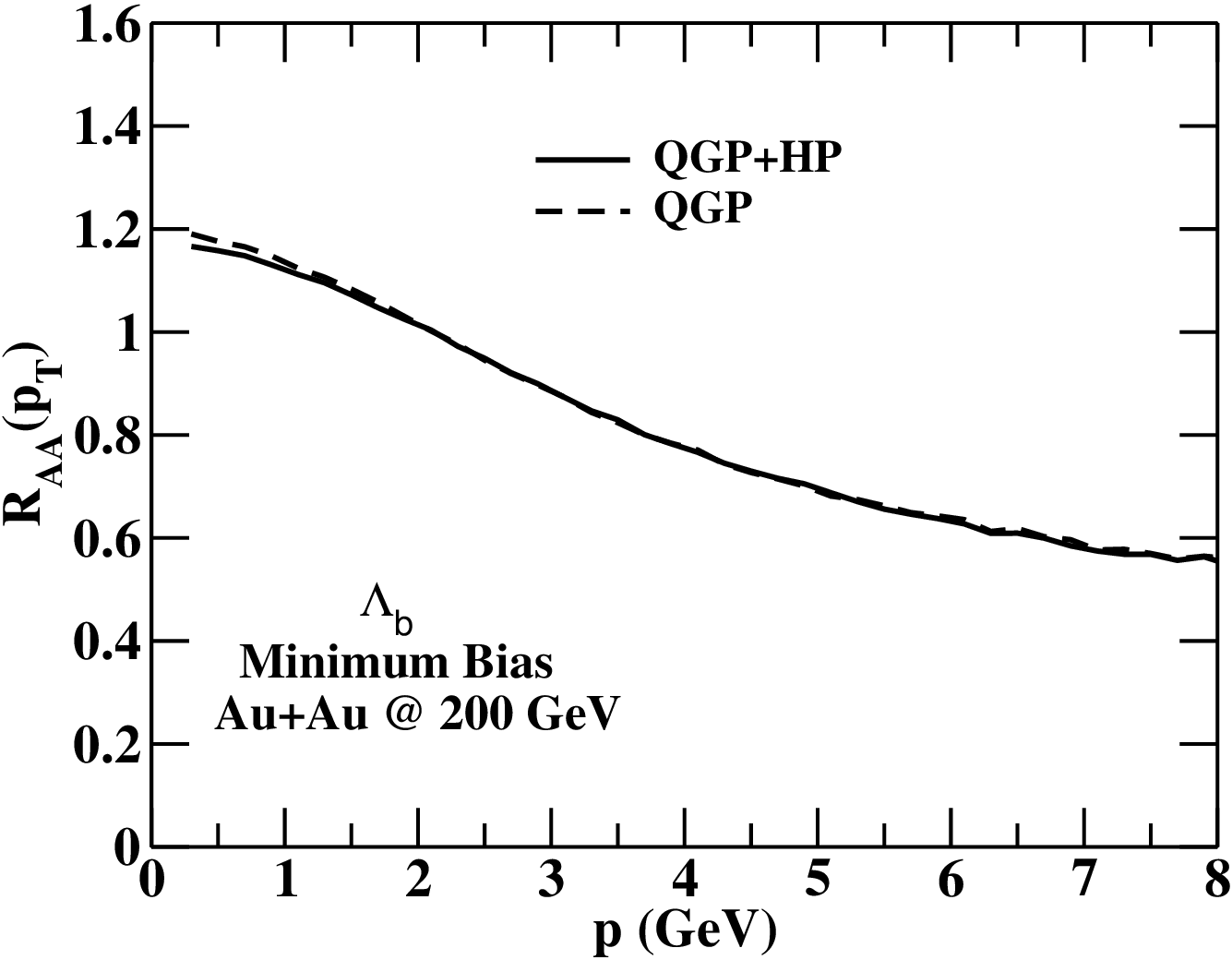}\hspace{2pc}
\caption{$R_{AA}$ as a function of $p_T$ for $\Lambda_c$ (left panel) and $\Lambda_b$  (right panel) at RHIC energy.}
\label{fig2}
\end{center}
\end{figure}

In Fig.~\ref{fig2} we show the variation of $R_{AA}$ as a function $p_T$ for the $\Lambda_c$ (left panel) 
and  $\Lambda_b$ (right panel) in the QGP as well as in the QGP+HP at RHIC energy. 
For the $\Lambda_c$ the suppression is stronger as we increase $p_T$ than the $\Lambda_b$, mainly due to 
the different interaction of c and b quarks in the QGP phase.  We find that the 
role of the hadronic phase on both the $\Lambda_c$ and the $\Lambda_b$ $R_{AA}$ is almost unnoticeable. 
This can be explained because $R_{AA}$ is very sensitive to the early stages of the 
expansion (at high temperatures) where the energy density
is the highest~\cite{Das:2015hla}. Therefore, collisions take place at a high rate in the early stages, before hadronization. 
This translates into a strong initial suppression ($R_{AA}$) which  then gets saturated within 3-4 fm due 
to the radial flow that is able to compensate the baryon energy loss. 
Hence, further rescattering in the hadronic medium is unable to alter this spectrum.

Note that the spectra of $\Lambda_c$ and $\Lambda_b$ baryons is obtained here from the fragmentation of
high-energy charm and bottom quarks using the Peterson fragmentation function. 
Such mechanism of hadronization may not be valid for low-momentum hadrons which are expected to be produced 
from the coalescence of a heavy quark with thermal light partons~\cite{Greco:2003vf}.

\begin{figure}[ht]
\begin{center}
\includegraphics[width=17pc,clip=true]{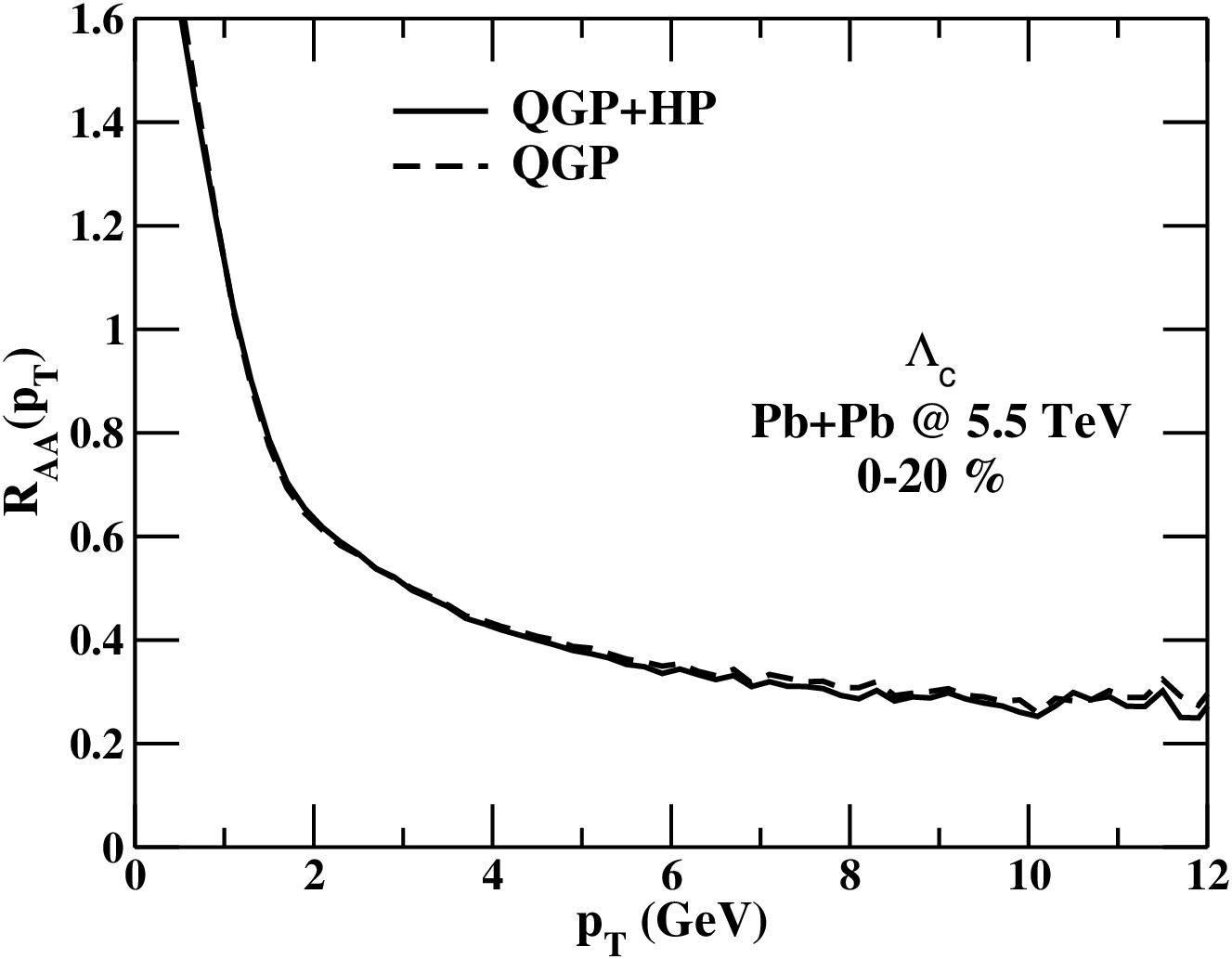}\hspace{2pc}
\includegraphics[width=17pc,clip=true]{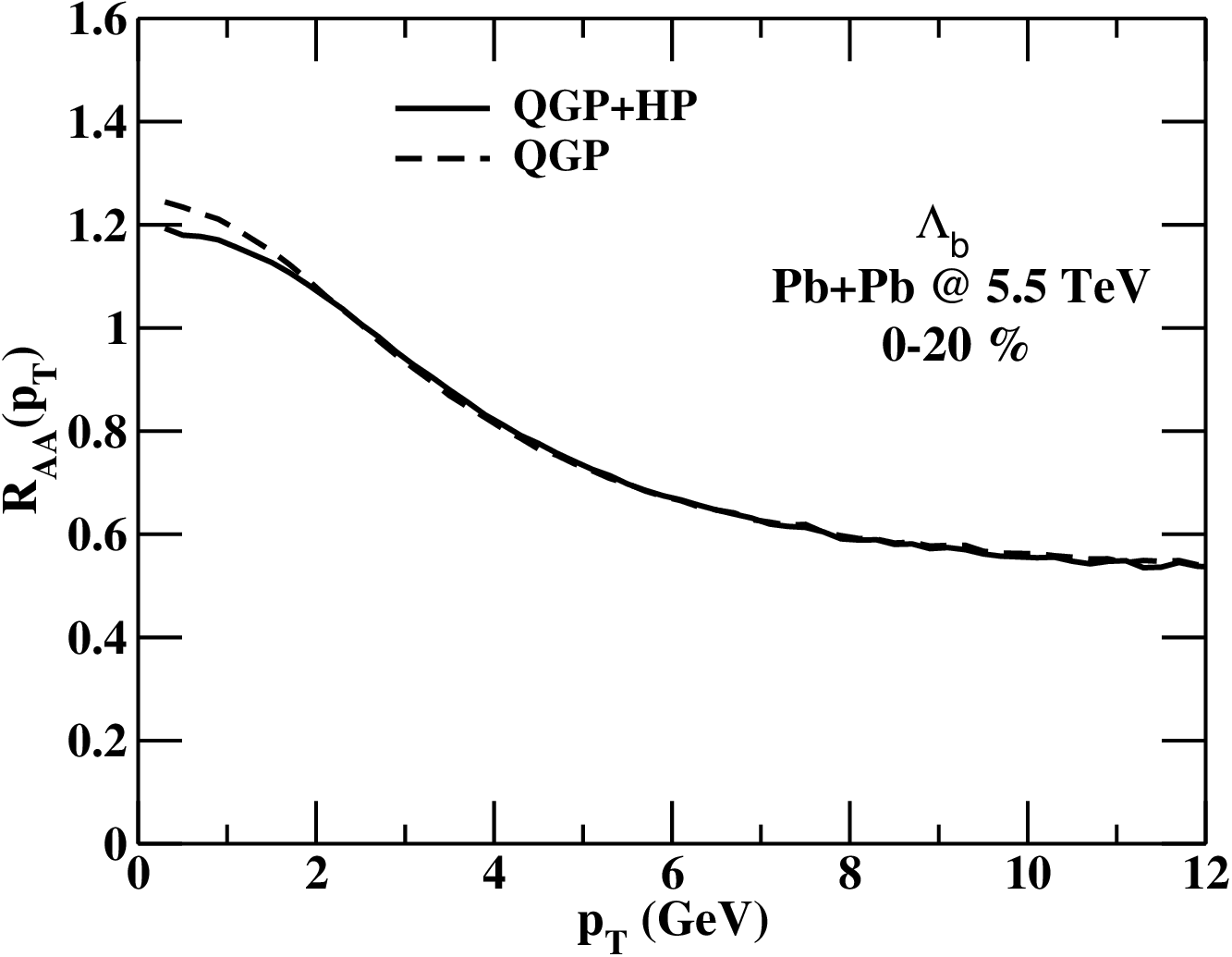}\hspace{2pc}
\caption{$R_{AA}$ as a function of $p_T$ for $\Lambda_c$ (left panel) and $\Lambda_b$  (right panel) at LHC energy.}
\label{fig3}
\end{center}
\end{figure}

We have also extended our calculation to study $R_{AA}$ of $\Lambda_c$ and $\Lambda_b$ at LHC colliding energy 
by performing simulations of $Pb+Pb$ at $\sqrt{s}$ = 5.5 ATeV.
These are our predictions for the upcoming heavy-baryon data at ALICE energy. In Fig.~\ref{fig3} we 
present the variation of $R_{AA}$ as a function of $p_T$ 
for the $\Lambda_c$  (left panel) and $\Lambda_b$ (right panel) in the QGP as well as in the QGP+HP at LHC energy. 
As seen before for RHIC energies, the suppression is stronger for $\Lambda_c$ than $\Lambda_b$. 
This is mainly due to the larger drag coefficient of charm quark than bottom quark, which shifts the high-$p_T$ particles 
to lower $p_T$ resulting in a higher population at low $p_T$. In addition, the bottom quark initial distribution is harder 
than the initial charm quark distribution. We have not considered the effect of shadowing~\cite{Eskola:2008ca} 
in the initial charm distribution which could be significant at low momentum. 
In the case of LHC energy we find that the role of the hadronic phase on $R_{AA}$ is almost unnoticeable for both $\Lambda_c$ and $\Lambda_b$. 
 
\subsection{Elliptic flow, $v_2$~\label{sec:v2}}

Another key observable related to heavy quarks measured at the RHIC and LHC energies is the elliptic flow induced by the spatial 
anisotropy of the bulk medium. It can be calculated as

\be
 v_2=\left\langle  \frac{p_x^2 -p_y^2}{p_T^2}\right\rangle =\left\langle  \frac{p_x^2 -p_y^2}{p_x^2+p_y^2}\right\rangle\ . \qquad \qquad
\ee
We define the $v_2$ generated in QGP phase taking $p_x$, $p_y$ and $p_T$ as the momenta of the 
heavy baryons  at $T_c$. 
The $v_2$ for the heavy baryons during the entire evolution process, from the 
beginning of the QGP phase to the end of the hadronic phase, is computed by 
taking $p_x$, $p_y$ and $p_T$ the momenta at the freeze-out $T_f$.

\begin{figure}[ht]
\begin{center}
\includegraphics[width=17pc,clip=true]{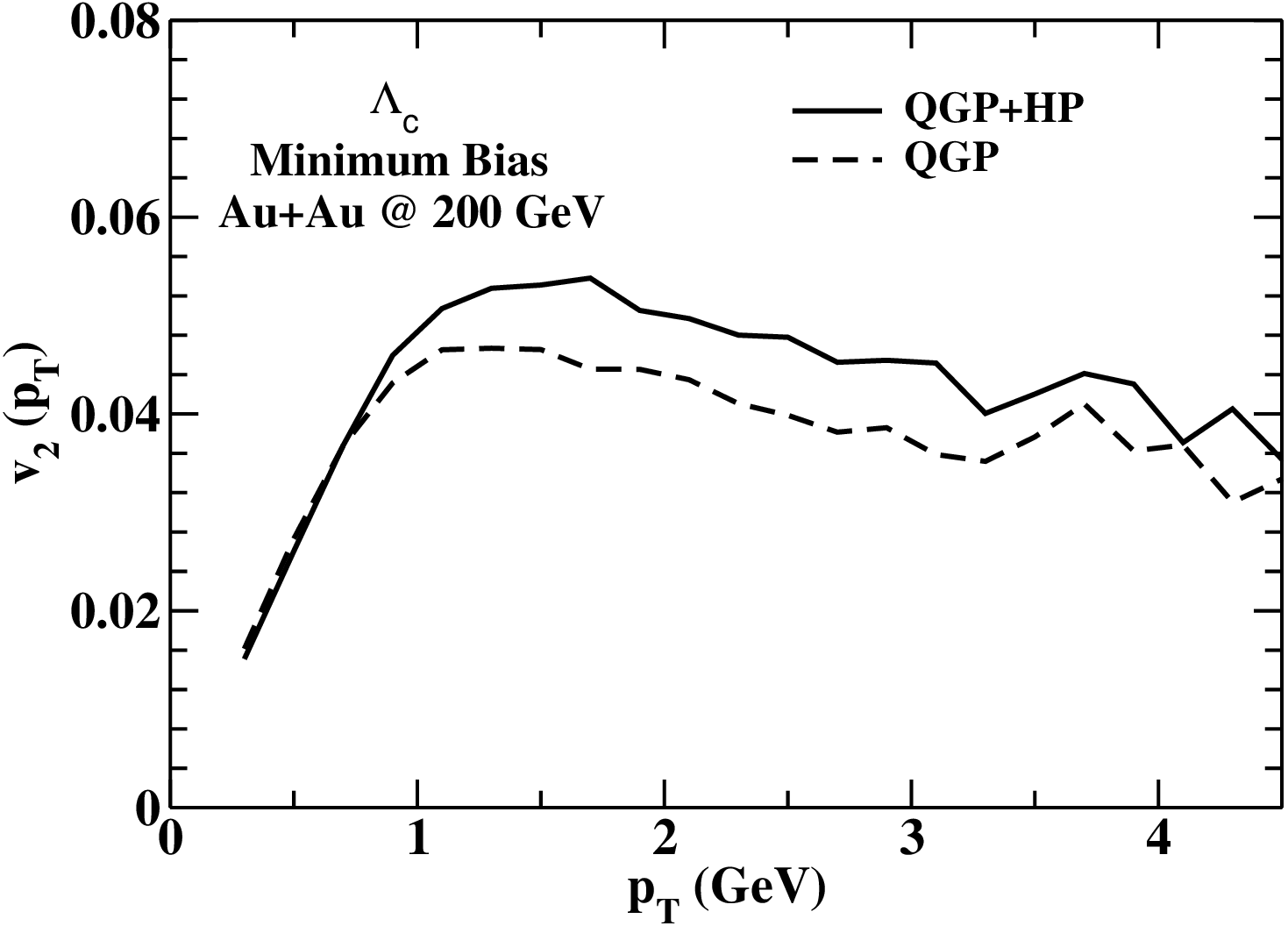}\hspace{2pc}
\includegraphics[width=17pc,clip=true]{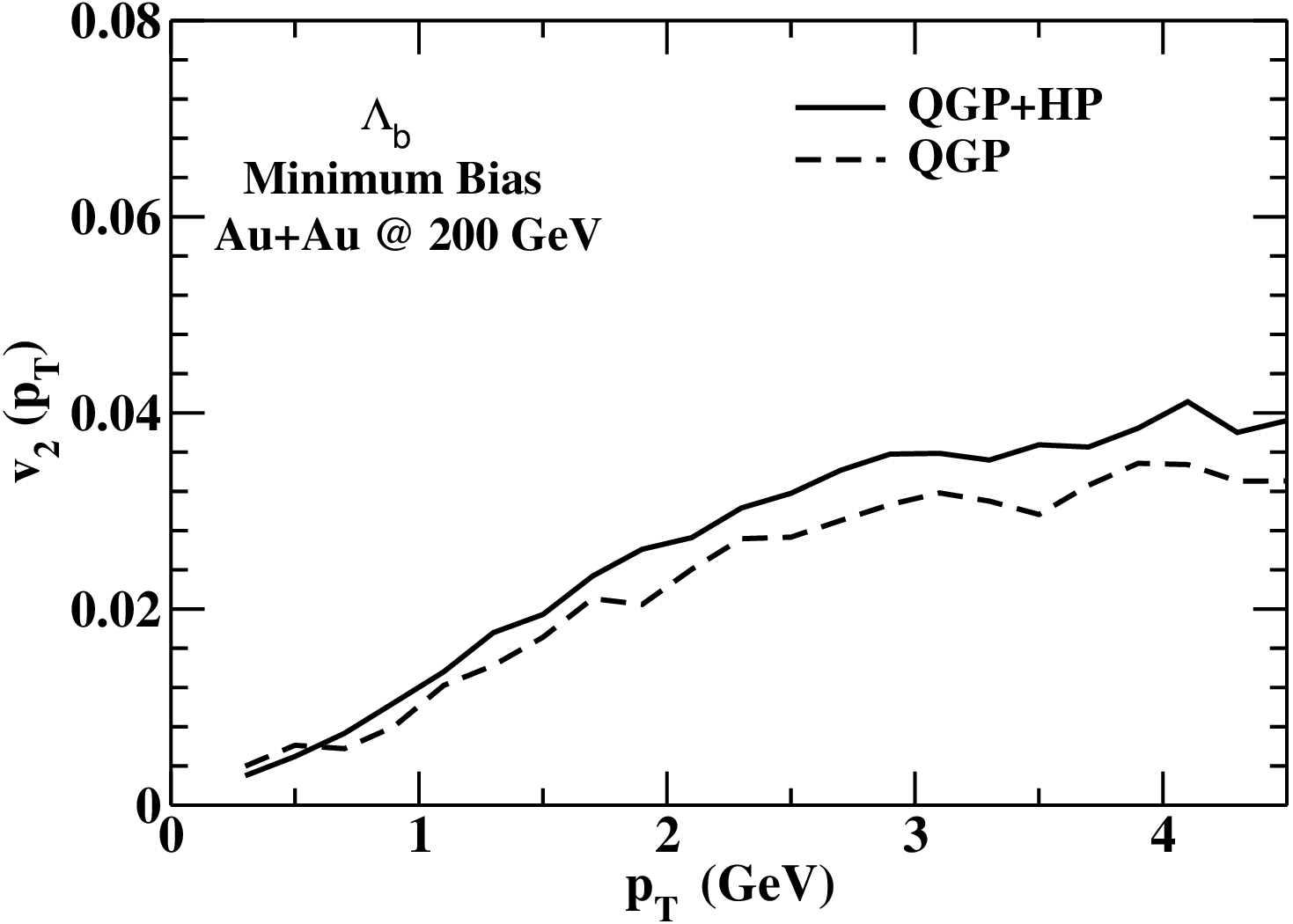}\hspace{2pc}
\caption{$v_2$ as a function of $p_T$ for $\Lambda_c$ (left panel) and $\Lambda_b$  (right panel) at RHIC energy.}
\label{fig4}
\end{center}
\end{figure}

In Fig.~\ref{fig4} we see the variation of $v_2$ as a function of $p_T$ for the $\Lambda_c$ (left panel) 
and  $\Lambda_b$ (right panel) in the QGP as well as in the QGP+HP at RHIC energy. 
We find that the $v_2$ is enhanced due to the presence of the hadronic phase. 
As mentioned earlier, the $R_{AA}$ is quite sensitive to the early stages of the expansion 
(at high $T$) where the energy density is the highest, and therefore
collisions take place at a higher rate. However such a strong interaction will not be accompanied 
by a build-up of $v_2$ because the bulk medium has not yet developed
a sizable part of its elliptic flow. First, the bulk will generate its own $v_2$ and then the 
bulk will transfer it to the heavy quarks. This usually happens at the later
stage of the evolution. Hence, the $v_2$ is sensitive to the heavy particle-bulk interaction.

It should be mentioned that the heavy baryons develop a substantial part 
of their $v_2$ mainly from the interaction they suffer at the quark level (as $c$ or $b$ quarks) 
in the QGP phase as well as due to their interaction in the hadronic phase. But they also can 
get some part of their $v_2$ (mainly at low momentum) from the 
thermal light quarks during hadronization by coalescence, which cannot be captured using only 
fragmentation.

\begin{figure}[ht]
\begin{center}
\includegraphics[width=17pc,clip=true]{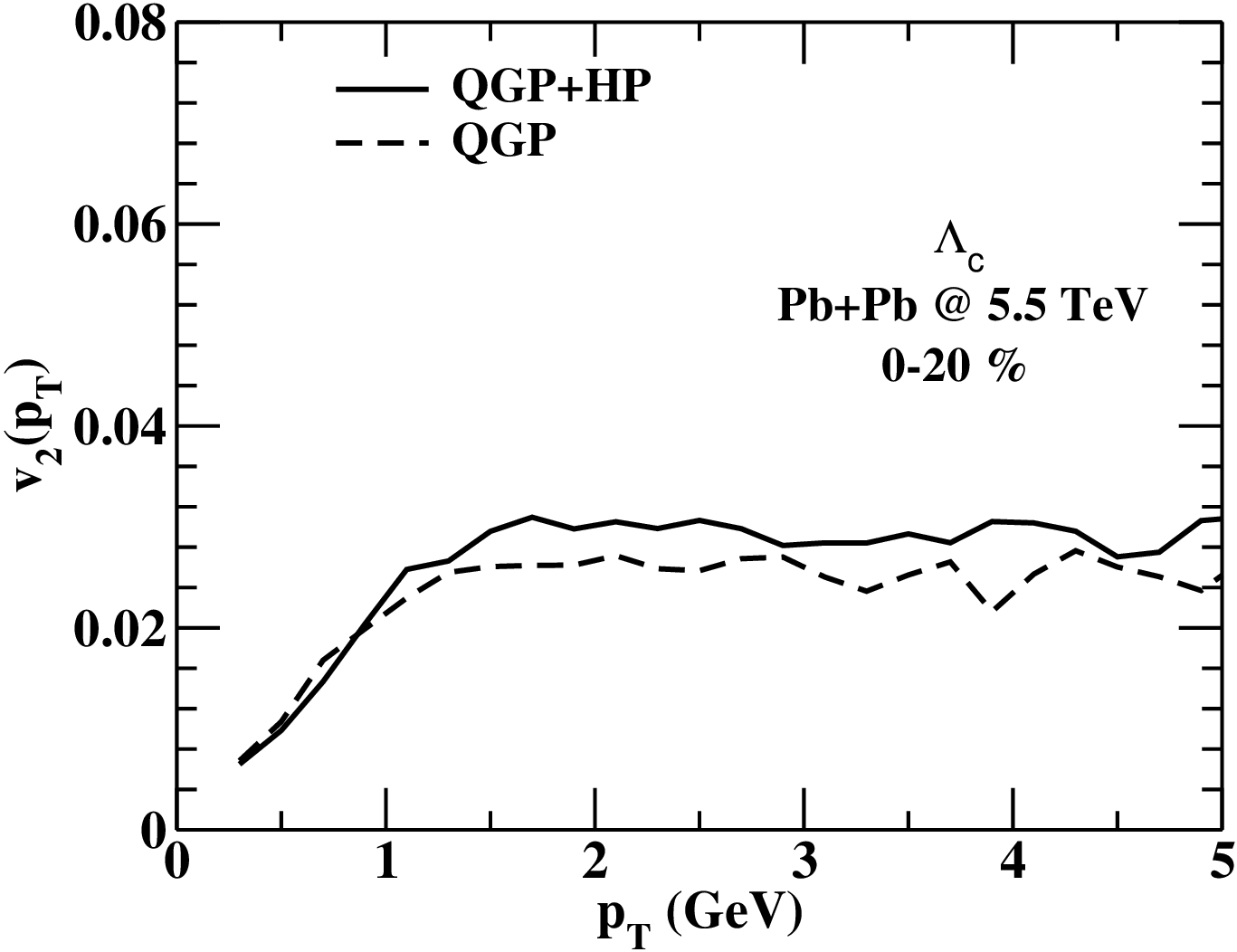}\hspace{2pc}
\includegraphics[width=17pc,clip=true]{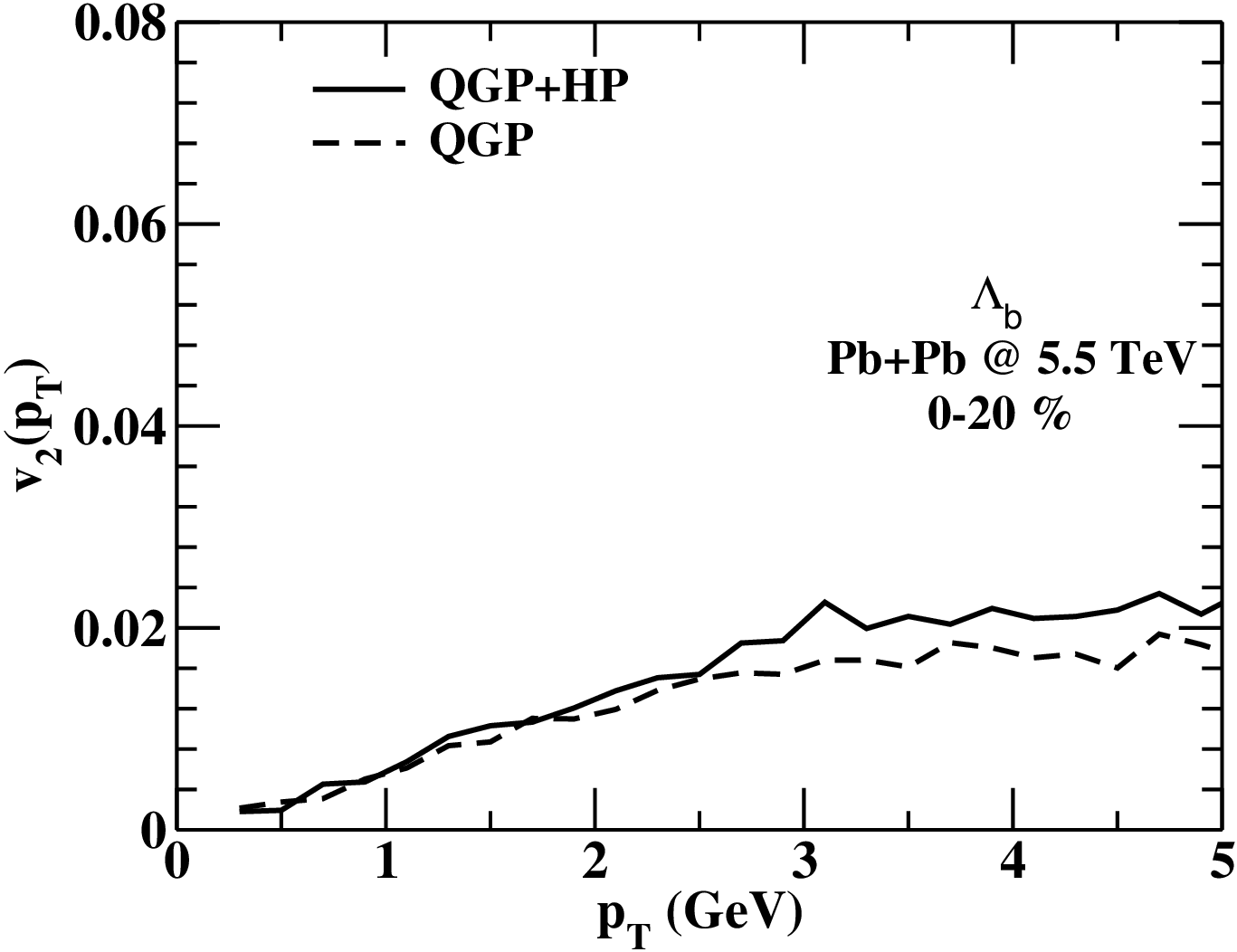}\hspace{2pc}
\caption{$v_2$ as a function of $p_T$ for $\Lambda_c$ (left panel) and $\Lambda_b$  (right panel) at LHC energy.}
\label{fig5}
\end{center}
\end{figure}

We have also extended the calculation for LHC energies at Run 2. In Fig.~\ref{fig5} we have shown the variation of $v_2$ as a 
function of $p_T$ for the $\Lambda_c$ (left panel) and $\Lambda_b$ (right panel)  in the QGP as 
well as in the QGP+HP phase at $\sqrt{s} = 5.5$ ATeV. We find 
the $v_2$ further enhanced upto 15 $\%$  due to the presence of the hadronic phase.

We find the enhancement of the $v_2$ due to the presence of hadronic phase is larger for RHIC 
colliding energy than the LHC energy,  which is clearly shown on the $v_2$ plots. 
The difference in the magnitude of $v_2$ due to the hadronic phase contribution
at the RHIC and LHC energies can be understood from the magnitude of the drag and 
diffusion coefficients in the hadronic medium as well as from the initial distribution. 
The coefficients and the initial distribution are inputs in  Langevin dynamics at the beginning of the hadronic phase. 
The temperature of the hadronic medium for both the RHIC and LHC colliding energies varies from $T_c$ to $T_f$
(from 160 to 120 MeV), and therefore the values of the drag and diffusion
coefficients will not change much. However, the input initial distribution to the hadronic matter 
is harder at the LHC energy than at the RHIC energy, resulting in less $v_2$ at the LHC energy.
Also the lifetime of the hadronic phase remains the same for both RHIC and LHC energies, whereas the lifetime
of the QGP phase is longer at LHC energy than RHIC, hence, having the hadronic phase 
less impact at LHC energy. Indeed, the effect of the hadronic phase on the $v_2$ will be more significant
for low-energy nuclear collision due to the diminishing lifetime of the QGP phase.

Note that, as compared to $v_2$,  the impact of the hadronic phase in $R_{AA}$ is much smaller both at RHIC and LHC energy. 
Hence, the $R_{AA}$ may play a unique role in characterizing QGP phase.
It is also important to mention that the impact of the hadronic medium is almost mass independent, i.e. the 
impact of the hadronic medium in the $v_2$ of $\Lambda_b$ and $\Lambda_c$ is similar.

\section{Heavy baryon to meson ratio~\label{sec:ratio}}
The heavy baryon to heavy meson ratios, ($\Lambda_c/D$ and $\Lambda_b/B$), are fundamental for 
the understanding of in the medium hadronisation~\cite{Greco:2003vf} with respect to the light flavored baryon to meson ratio~\cite{Greco:2003xt,Greco:2003mm}.
Enhancement of $\Lambda_c/D$ and $\Lambda_b/B$ in Au+Au/Pb+Pb collisions compared to p+p collisions affects  
the non-photonic single-electron spectra resulting from semileptonic
decays of hadrons containing heavy flavors, hence, their  nuclear suppression factor ($R_{AA}$)~\cite{Sorensen:2005sm,MartinezGarcia:2007hf,Ayala:2009pe,Fries:2008hs}. 
This is because the branching ratio for the decay process $\Lambda_c \rightarrow e + X (4.5\% \pm 1.7\%)$ is 
smaller than the decay process $D \rightarrow e + X (17.2\% \pm 1.9\%)$, resulting in less electrons 
from decays of $\Lambda_c$ baryon than D meson. Hence, enhancement of $\Lambda_c/D$  
ratio in Au+Au/Pb+Pb collisions will affect the observed non-photonic single-electrons, hence, the $R_{AA}$.
In this manuscript we investigate the $\Lambda_c/D$ ratio for both RHIC and LHC energies. 
We study the possible impact of hadronic medium, if any, on 
heavy baryon to heavy meson ratio. This investigation is very timely, because LHC is preparing for Run 2 and 3 
having major interest on heavy baryon to meson ratio. It becomes particularly
appealing to study if heavy baryons observables are carrying signature of the QGP phase or QGP+Hadronic 
phase.

\begin{figure}[ht]
\begin{center}
\includegraphics[width=17pc,clip=true]{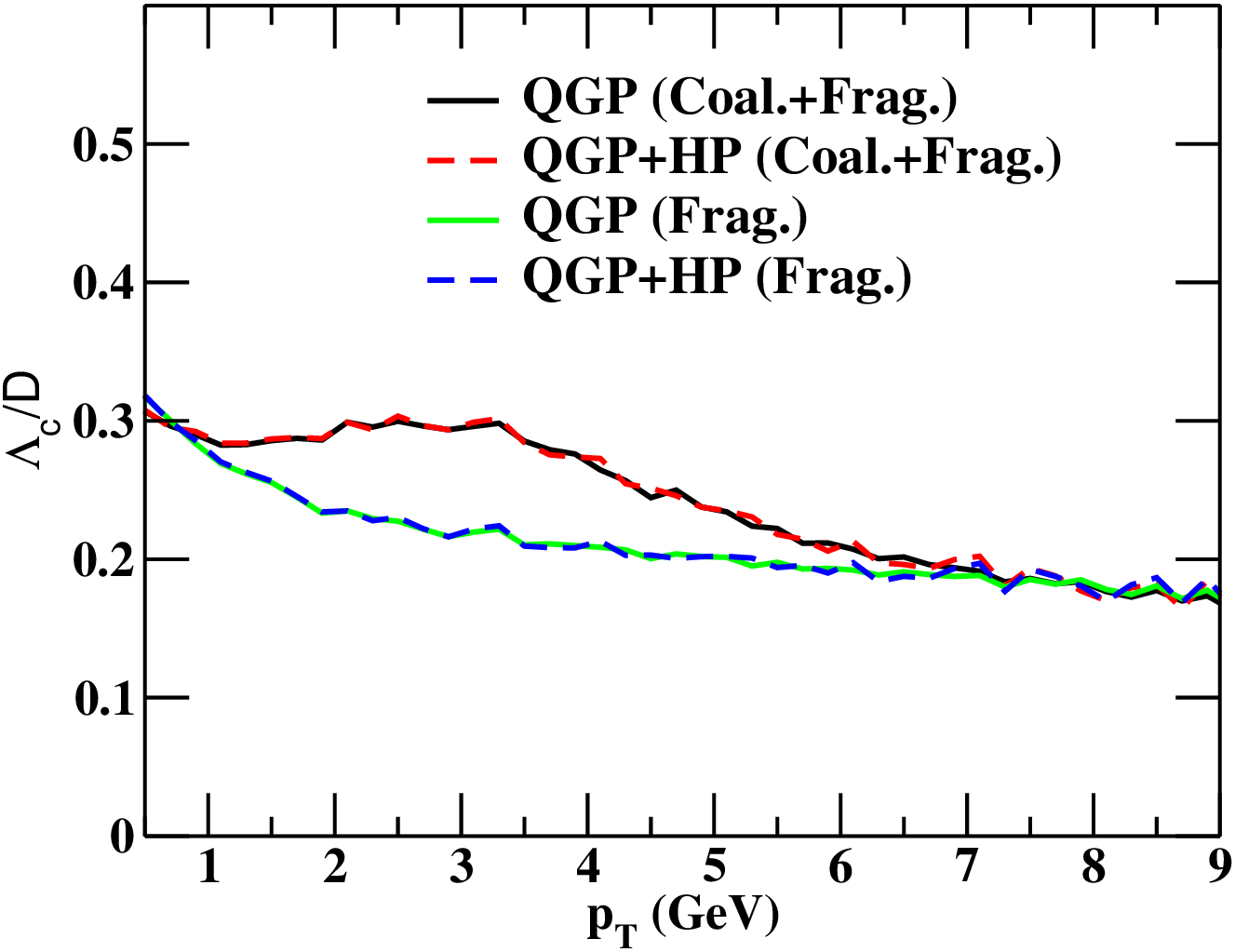}\hspace{2pc}
\includegraphics[width=17pc,clip=true]{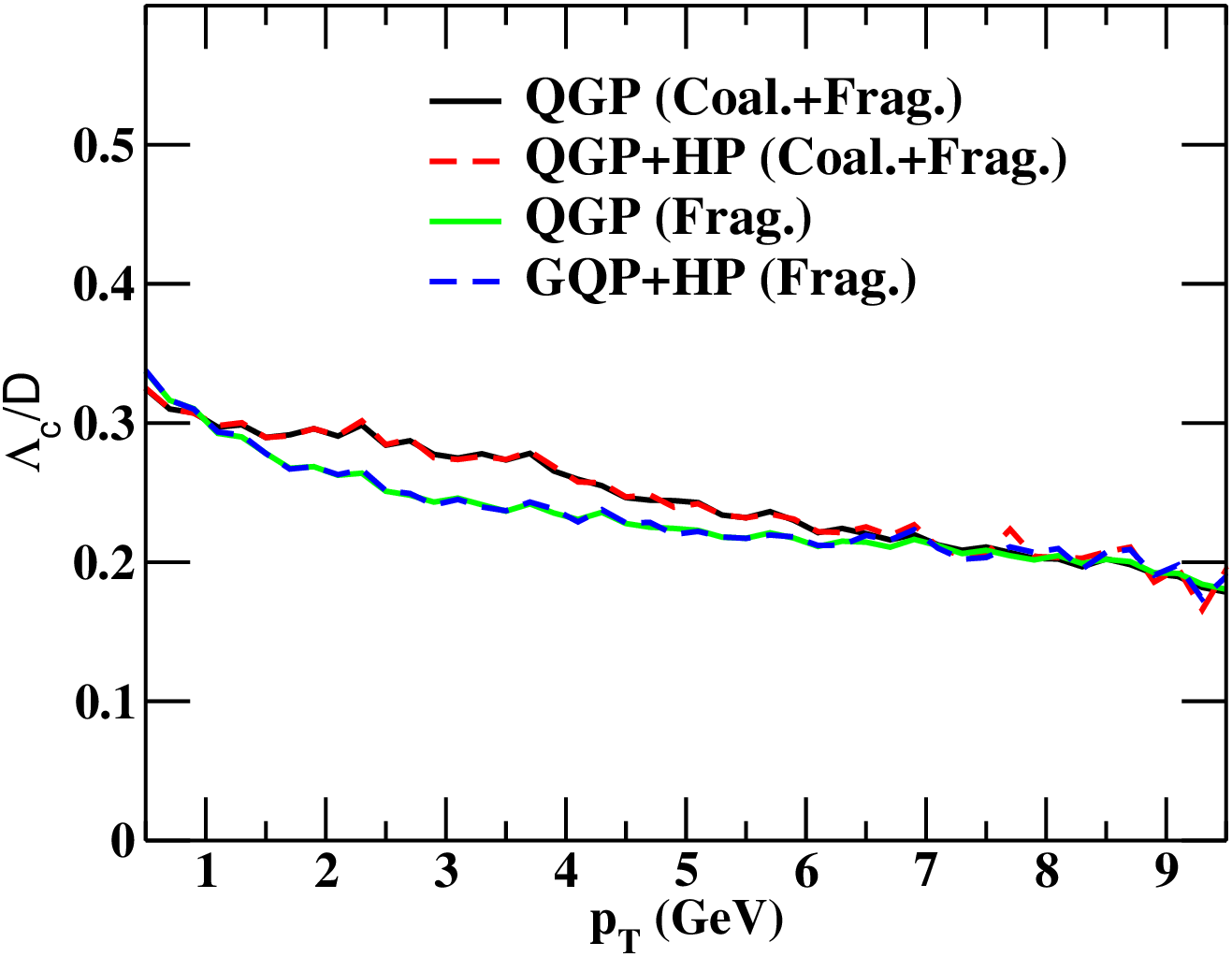}\hspace{2pc}
\caption{$\Lambda_c/D$ as a function of $p_T$ at RHIC (left panel) and at LHC energy (right panel).}
\label{fig6}
\end{center}
\end{figure}

To evaluate the heavy baryon to meson ratio we use the fragmentation as well as fragmentation plus coalescence model for heavy qurak hadronization.    
The coalescence mechanism we employ for D meson and $\Lambda_c$ is similar to one used for the 
hadronization of light quarks in \cite{Minissale:2015zwa,Greco:2003xt,Greco:2003mm}.
Given the momentum distribution of the heavy quarks obtained solving the Langevin dynamics, the contribution due to 
coalescence can be evaluated as follows:
\begin{eqnarray}
\label{eq-coal}
\frac{d^{2}N_{H}}{dP_{T}^{2}}&=& g_{H} \int \prod^{n}_{i=1} \frac{d^{3}p_{i}}{(2\pi)^{3}E_{i}} p_{i} \cdot d\sigma_{i}  \; f_{q_i}(x_{i}, p_{i})
 f_{H}(x_{1}..x_{n}, p_{1}..p_{n})\, \delta^{(2)} \left(P_{T}-\sum^{n}_{i=1} p_{T,i} \right)
\label{eq_coal}
\end{eqnarray}
where $d\sigma_{i}$ represents an element of a space-like hypersurface, $n$ is the number of quarks,  $g_{H}$ is a statistical 
factor to form a colorless hadron from the spin 1/2 quark and antiquark. $f_{q_i}$ are the quark/anti-quark distribution functions. 
$f_{H}$ is the Wigner function and describes the coordinate and momentum distribution of quarks/anti-quarks in a hadron. 
The Wigner function  depends in principle on the overlap of the quark and anti-quark distribution functions with the
wave function of the meson/baryons  as well as the interactions of emitted virtual partons, which are needed for the 
energy and momentum balance, with the QGP. If one neglecte the off-shell effects  then the  coalescence probability
function is simply the covariant hadron Wigner distribution function. The longitudinal momentum distributions of the quarks and 
antiquarks are assumed to be boost-invarian. For details, we refer to Ref.\cite{Greco:2003mm}. 

In the Greco-Ko-Levai (GKL)  approach~\cite{Greco:2003mm} for a heavy meson the Wigner function is taken as a Gaussian of 
radius $\Delta_{x}$ in the coordinate  and $\Delta_{p}$ in the 
momentum space,  these two parameters being  by the uncertainty principle $\Delta_{x}\Delta_{p}=1$,  

\begin{eqnarray} 
f_{M}(x_{1}, x_{2}; p_{1}, p_{2}) &=& 8
\exp(x_r^2/(2\Delta_x^2))\exp((p_r^2-\Delta m_{12}^{2})/(2\Delta_p^2))
\end{eqnarray}
where $x_{r}=x_{1} - x_{2}$ and $p_{r}=p_1-p_2$ are the quadri-vectors for the relative coordinates 
and  $\Delta m_{12}=m_1-m_2$ is the scalar
We use $\Delta_x$= 1.06 fm for D meson.

To extend the calculations for mesons to the formation of baryons  from the parton
distribution functions, we take the baryon coalescence probability function as,

\begin{eqnarray}
&&F_B(x_1,x_2,x_3;p_1,p_2,p_3) \nonumber \\
&&=8^2\exp(x_r^2/(2\Delta_x^2))\exp((p_r^2-\Delta m_{12}^{2})/(2\Delta_p^2)) \nonumber\\
&&\times \exp(\frac{1}{6}(x_1+x_2-2x_3)^2/(2\Delta_x^2))  \nonumber \\ 
&&\times \exp(\frac{1}{6}((p_1+p_2-2p_3)^2-(m_1+m_2-2m_3)^2)/(2\Delta_p^2))  
\end{eqnarray} 
We use $\Delta_x$= 0.98 fm for $\Lambda_c$.
Starting from the charm quark distributions, coalescence probability of  $D$ and $\Lambda_c$ has been calculated using Eq.~\ref{eq-coal} at $T_c$ 
with the appropriate choice of Wigner function.
The charm quarks that do not coalescence, are eventually fragmented in accordance with the fragmentation functions of $D$ and $\Lambda_c$ 
discussed in subsection ~\ref{sec:frag}. In the present calculations, we have included the contributions from resonances decay coming from $\Sigma_c$, 
$\Lambda(2526)$, $\bar \Sigma_c$ and $\bar \Lambda_c$. It can be mentioned that the contribution from resonance decays affect the ratio (heavy baryon to meson)~\cite{Oh:2009zj}
as it involve the ratio of two different hadron species having different contribution from the resonance decays. But the impact of resonances 
decays on $R_{AA}$ is negligible, even if not vanishing, as it affects similarly the numerator and denominator of the ratio because its impact is similar
in p+p and Au+Au/Pb+Pb. After the hadronization, we perform the time evolution of heavy hadrons ($D$ and $\Lambda_c$) within the 
hadronic phase until the temperature reaches $T_{kin}=120$ MeV.
 
In Fig.~\ref{fig6} we see the variation of $\Lambda_c/D$  as a function of $p_T$ at RHIC (left panel) and  at LHC energy (right panel) for QGP 
and QGP+Hadronic phase within coalescence plus fragmentation and fragmentation. The impact of coalescence is mainly restricted to the 
low $p_T$ region (within $p_T$~1-5 GeV) above which the hadronization mechanism is dominated by fragmentation. The coalescence probability involves the 
product of two distribution functions which fall very fast at high momentum making way for the fragmentation as the dominant mechanism of hadronization.
As shown, the impact of coalescence is less significant at LHC energy than RHIC energy. This is because impact of coalescence depends on the slope of the charm 
quark $p_T$ distribution. For a harder distribution the gain in momentum reflects in a smaller increase of the slope of the charm quark distribution, instead 
if the distribution decreases fast in momentum then the same momentum gain due to coalescence will result in a stronger increase of the spectrum. 
For a hard distribution, which is the case at LHC energy (in contrast to RHIC),the impact of coalescence will be less pronounce. 
The present study gives the possibility to disentangle/understand different hadronization mechanisms of heavy quarks once the data will be available.
More significantly, the $\Lambda_c/D$ is independent of the hadronic phase both at RHIC and LHC energies. As discussed earlier, the impact of hadronic 
phase is almost unnoticeable on  $R_{AA}$ of $\Lambda_c$, hence, on the spectra. 
Thus,  the impact of the hadronic medium on another ratio, such as $\Lambda_c/D$,  is negligible, which enables $\Lambda_c/D$ as a noble probe of QGP phase
dynamics including its hadronization. We prefer to ignore 
the bottom case to avoid uncertainty in the final ratios due to lack of knowledge of the resonance feed-down from higher states. 

\section{Conclusions~\label{sec:conclusions}}
We have studied the $R_{AA}$ and $v_2$ of heavy baryons highlighting the impact of the hadronic medium 
on these observables  within a Langevin dynamics. The QGP medium interaction of the $c$ and $b$ quarks with 
the light quarks and gluons have been treated within a quasi-particle model~\cite{Das:2015ana}. 
To fix the shape of heavy baryon fragmentation function, we have used the information available 
from electron-positron annihilation data on heavy baryon fragmentation function. 
Heavy baryon fragmentation function is softer than the heavy meson fragmentation function as 
the baryon involves three-body fragmentation whereas the meson a two-body fragmentation. We find that the impact of
hadronic medium on the $R_{AA}$ for heavy baryons ($\Lambda_c$ and $\Lambda_b$) is almost unnoticeable while 
the hadronic medium contribution is sizable on  $v_2$, which is about $20\%$. We have also calculated the 
$\Lambda_c/D$ ratio at RHIC and LHC energies. The heavy hadron suppression does not 
change in the hadronic phase, hence the spectra. Thus,  the impact of the hadronic medium on another 
ratio, such as $\Lambda_c/D$, is negligible, which enables $\Lambda_c/D$ as a noble probe of QGP phase. 
So the enhancement of heavy baryon to meson ratio, if any, 
in $Au+Au$/$Pb+Pb$ collisions with respect to $p+p$ collisions would be an indication of QGP phase dynamics 
including its hadronization. Furthermore, the $\Lambda_c/D$ can serve as a tool to disentangle different 
hadronization mechanisms once the data will be available.

\section{Acknowledgements}
JMTR thanks Jan Wagner for interesting discussions on heavy-quark physics.
SKD, FS and VG acknowledge the support by the ERC StG under the QGPDyn Grant no. 25968.
JMTR acknowledges the financial support from programme TOGETHER from R\'egion Pays de la Loire, 
and from a Helmholtz Young Investigator Group VH-NG-822 from the Helmholtz Association and GSI.
LT acknowledges support from the Ram\'on y Cajal research programme.
JMTR and LT also acknowledge support by Grants FPA2010-16963 and FPA2013-43425-P (Spain).


\begin{thebibliography}{9}

\bibitem{Prino:2016cni} 
  F.~Prino and R.~Rapp,
  J.\ Phys.\ G {\bf 43}, no. 9, 093002 (2016)
  doi:10.1088/0954-3899/43/9/093002
  [arXiv:1603.00529 [nucl-ex]].
  
  
\bibitem{Adamczyk:2014uip} 
  L.~Adamczyk {\it et al.} [STAR Collaboration],
  Phys.\ Rev.\ Lett.\  {\bf 113}, no. 14, 142301 (2014)
  doi:10.1103/PhysRevLett.113.142301
  [arXiv:1404.6185 [nucl-ex]].


\bibitem{ALICE:2012ab} 
  B.~Abelev {\it et al.} [ALICE Collaboration],
  JHEP {\bf 1209}, 112 (2012)
  doi:10.1007/JHEP09(2012)112
  [arXiv:1203.2160 [nucl-ex]].


\bibitem{Abelev:2013lca} 
  B.~Abelev {\it et al.} [ALICE Collaboration],
  Phys.\ Rev.\ Lett.\  {\bf 111}, 102301 (2013)
  doi:10.1103/PhysRevLett.111.102301
  [arXiv:1305.2707 [nucl-ex]].

\bibitem{Abelev:2014ipa} 
  B.~B.~Abelev {\it et al.} [ALICE Collaboration],
  Phys.\ Rev.\ C {\bf 90},  034904 (2014)

 
\bibitem{Adam:2015sza} 
  J.~Adam {\it et al.} [ALICE Collaboration],
  JHEP {\bf 1603}, 081 (2016)
  doi:10.1007/JHEP03(2016)081
  [arXiv:1509.06888 [nucl-ex]].


\bibitem{Linnyk:2008hp} 
  O.~Linnyk, E.~L.~Bratkovskaya and W.~Cassing,
  Int.\ J.\ Mod.\ Phys.\ E {\bf 17}, 1367 (2008)
  doi:10.1142/S0218301308010507
  [arXiv:0808.1504 [nucl-th]].
  

\bibitem{Das:2010tj} 
  S.~K.~Das, J.~e.~Alam and P.~Mohanty,
  Phys.\ Rev.\ C {\bf 82}, 014908 (2010)
  doi:10.1103/PhysRevC.82.014908
  [arXiv:1003.5508 [nucl-th]]; arXiv:0908.4194 [nucl-th].

\bibitem{Mazumder:2011nj} 
  S.~Mazumder, T.~Bhattacharyya, J.~e.~Alam and S.~K.~Das,
  Phys.\ Rev.\ C {\bf 84}, 044901 (2011)
  doi:10.1103/PhysRevC.84.044901
  [arXiv:1106.2615 [nucl-th]].
  
\bibitem{Lang:2012yf} 
  T.~Lang, H.~van Hees, J.~Steinheimer and M.~Bleicher,
  arXiv:1208.1643 [hep-ph].
  
\bibitem{He:2012df} 
  M.~He, R.~J.~Fries and R.~Rapp,
  Phys.\ Rev.\ Lett.\  {\bf 110}, no. 11, 112301 (2013)
  doi:10.1103/PhysRevLett.110.112301
  [arXiv:1204.4442 [nucl-th]].
  
\bibitem{He:2011qa} 
  M.~He, R.~J.~Fries and R.~Rapp,
  Phys.\ Rev.\ C {\bf 86}, 014903 (2012)
  doi:10.1103/PhysRevC.86.014903
  [arXiv:1106.6006 [nucl-th]].
   
\bibitem{Das:2013kea} 
  S.~K.~Das, F.~Scardina, S.~Plumari and V.~Greco,
  Phys.\ Rev.\ C {\bf 90}, 044901 (2014)
  doi:10.1103/PhysRevC.90.044901
  [arXiv:1312.6857 [nucl-th]].

\bibitem{Alberico:2011zy} 
  W.~M.~Alberico, A.~Beraudo, A.~De Pace, A.~Molinari, M.~Monteno, M.~Nardi and F.~Prino,
  Eur.\ Phys.\ J.\ C {\bf 71}, 1666 (2011)
  doi:10.1140/epjc/s10052-011-1666-6
  [arXiv:1101.6008 [hep-ph]].
  
\bibitem{Uphoff:2011ad} 
  J.~Uphoff, O.~Fochler, Z.~Xu and C.~Greiner,
  Phys.\ Rev.\ C {\bf 84}, 024908 (2011)
  doi:10.1103/PhysRevC.84.024908
  [arXiv:1104.2295 [hep-ph]].
  
\bibitem{Uphoff:2012gb} 
  J.~Uphoff, O.~Fochler, Z.~Xu and C.~Greiner,
  Phys.\ Lett.\ B {\bf 717}, 430 (2012)
  doi:10.1016/j.physletb.2012.09.069
  [arXiv:1205.4945 [hep-ph]].

\bibitem{Cao:2013ita} 
  S.~Cao, G.~Y.~Qin and S.~A.~Bass,
  Phys.\ Rev.\ C {\bf 88}, 044907 (2013)
  doi:10.1103/PhysRevC.88.044907
  [arXiv:1308.0617 [nucl-th]].
  
\bibitem{Cao:2015hia} 
  S.~Cao, G.~Y.~Qin and S.~A.~Bass,
  Phys.\ Rev.\ C {\bf 92}, no. 2, 024907 (2015)
  doi:10.1103/PhysRevC.92.024907
  [arXiv:1505.01413 [nucl-th]].


\bibitem{Gossiaux:2008jv} 
  P.~B.~Gossiaux and J.~Aichelin,
  Phys.\ Rev.\ C {\bf 78}, 014904 (2008)
  doi:10.1103/PhysRevC.78.014904
  [arXiv:0802.2525 [hep-ph]].
  
  

\bibitem{Nahrgang:2013xaa} 
  M.~Nahrgang, J.~Aichelin, P.~B.~Gossiaux and K.~Werner,
  Phys.\ Rev.\ C {\bf 89}, no. 1, 014905 (2014)
  doi:10.1103/PhysRevC.89.014905
  [arXiv:1305.6544 [hep-ph]].

\bibitem{Nahrgang:2014vza} 
  M.~Nahrgang, J.~Aichelin, S.~Bass, P.~B.~Gossiaux and K.~Werner,
  Phys.\ Rev.\ C {\bf 91}, no. 1, 014904 (2015)
  doi:10.1103/PhysRevC.91.014904
  [arXiv:1410.5396 [hep-ph]].


\bibitem{Ozvenchuk:2014rpa} 
  V.~Ozvenchuk, J.~M.~Torres-Rincon, P.~B.~Gossiaux, L.~Tolos and J.~Aichelin,
  Phys.\ Rev.\ C {\bf 90}, no. 5, 054909 (2014)

  
\bibitem{Song:2015sfa} 
  T.~Song, H.~Berrehrah, D.~Cabrera, J.~M.~Torres-Rincon, L.~Tolos, W.~Cassing and E.~Bratkovskaya,
  Phys.\ Rev.\ C {\bf 92}, no. 1, 014910 (2015)
  doi:10.1103/PhysRevC.92.014910
  [arXiv:1503.03039 [nucl-th]].


\bibitem{Song:2015ykw} 
  T.~Song, H.~Berrehrah, D.~Cabrera, W.~Cassing and E.~Bratkovskaya,
  Phys.\ Rev.\ C {\bf 93}, no. 3, 034906 (2016)
  doi:10.1103/PhysRevC.93.034906
  [arXiv:1512.00891 [nucl-th]].
  
  
  
\bibitem{Abelevetal:2014dna} 
  B.~Abelev {\it et al.} [ALICE Collaboration],
  J.\ Phys.\ G {\bf 41}, 087002 (2014)


\bibitem{Andronic:2015wma} 
  A.~Andronic {\it et al.},
  Eur.\ Phys.\ J.\ C {\bf 76}, no. 3, 107 (2016)
  doi:10.1140/epjc/s10052-015-3819-5
  [arXiv:1506.03981 [nucl-ex]].

  
\bibitem{Dainese:2016dea} 
  A.~Dainese {\it et al.},
  Frascati Phys.\ Ser.\  {\bf 62} (2016)
  [arXiv:1602.04120 [nucl-ex]].

\bibitem{Aaij:2013mga} 
  R.~Aaij {\it et al.} [LHCb Collaboration],
  Nucl.\ Phys.\ B {\bf 871}, 1 (2013)
  
\bibitem{Oh:2009zj} 
  Y.~Oh, C.~M.~Ko, S.~H.~Lee and S.~Yasui,
  Phys.\ Rev.\ C {\bf 79}, 044905 (2009)
  doi:10.1103/PhysRevC.79.044905
  [arXiv:0901.1382 [nucl-th]].
  
\bibitem{Lee:2007wr} 
  S.~H.~Lee, K.~Ohnishi, S.~Yasui, I.~K.~Yoo and C.~M.~Ko,
  Phys.\ Rev.\ Lett.\  {\bf 100}, 222301 (2008)
  doi:10.1103/PhysRevLett.100.222301
  [arXiv:0709.3637 [nucl-th]].
  
\bibitem{Sorensen:2005sm} 
  P.~R.~Sorensen and X.~Dong,
  Phys.\ Rev.\ C {\bf 74}, 024902 (2006)
  doi:10.1103/PhysRevC.74.024902
  [nucl-th/0512042].

\bibitem{Ayala:2009pe} 
  A.~Ayala, J.~Magnin, L.~M.~Montano and G.~T.~Sanchez,
  Phys.\ Rev.\ C {\bf 80}, 064905 (2009)
  doi:10.1103/PhysRevC.80.064905
  [arXiv:0908.0361 [nucl-th]].
  

\bibitem{MartinezGarcia:2007hf} 
  G.~Martinez-Garcia, S.~Gadrat and P.~Crochet,
  Phys.\ Lett.\ B {\bf 663}, 55 (2008)
  [Phys.\ Lett.\ B {\bf 666}, 533 (2008)]
  doi:10.1016/j.physletb.2008.07.061, 10.1016/j.physletb.2008.01.079
  [arXiv:0710.2152 [hep-ph]].
  
\bibitem{Greco:2003vf} 
  V.~Greco, C.~M.~Ko and R.~Rapp,
  Phys.\ Lett.\ B {\bf 595}, 202 (2004)
  doi:10.1016/j.physletb.2004.06.064
  [nucl-th/0312100].
  
  
\bibitem{Fries:2008hs} 
  R.~J.~Fries, V.~Greco and P.~Sorensen,
  Ann.\ Rev.\ Nucl.\ Part.\ Sci.\  {\bf 58}, 177 (2008)
  doi:10.1146/annurev.nucl.58.110707.171134
  [arXiv:0807.4939 [nucl-th]].
  
  
\bibitem{Tolos:2016slr} 
  L.~Tolos, J.~M.~Torres-Rincon and S.~K.~Das,
  Phys.\ Rev.\ D {\bf 94}, no. 3, 034018 (2016)
  doi:10.1103/PhysRevD.94.034018
  [arXiv:1601.03743 [hep-ph]].
  
\bibitem{Ghosh:2014oia} 
  S.~Ghosh, S.~K.~Das, V.~Greco, S.~Sarkar and J.~e.~Alam,
  Phys.\ Rev.\ D {\bf 90},  054018 (2014)
  
\bibitem{Das:2015ana} 
  S.~K.~Das, F.~Scardina, S.~Plumari and V.~Greco,
  Phys.\ Lett.\ B {\bf 747}, 260 (2015)


\bibitem{Moore:2004tg} 
  G.~D.~Moore and D.~Teaney,
  Phys.\ Rev.\ C {\bf 71}, 064904 (2005)
  doi:10.1103/PhysRevC.71.064904
  [hep-ph/0412346].

\bibitem{vanHees:2005wb} 
  H.~van Hees, V.~Greco and R.~Rapp,
  Phys.\ Rev.\ C {\bf 73}, 034913 (2006)
  doi:10.1103/PhysRevC.73.034913
  [nucl-th/0508055].
  
  
\bibitem{Cao:2011et} 
  S.~Cao and S.~A.~Bass,
  Phys.\ Rev.\ C {\bf 84}, 064902 (2011)
  doi:10.1103/PhysRevC.84.064902
  [arXiv:1108.5101 [nucl-th]].
  
  
\bibitem{Das:2012ck} 
  S.~K.~Das, V.~Chandra and J.~e.~Alam,
  J.\ Phys.\ G {\bf 41}, 015102 (2013)
  doi:10.1088/0954-3899/41/1/015102
  [arXiv:1210.3905 [nucl-th]].


\bibitem{Berrehrah:2013mua} 
  H.~Berrehrah, E.~Bratkovskaya, W.~Cassing, P.~B.~Gossiaux, J.~Aichelin and M.~Bleicher,
  Phys.\ Rev.\ C {\bf 89}, no. 5, 054901 (2014)
  doi:10.1103/PhysRevC.89.054901
  [arXiv:1308.5148 [hep-ph]].


\bibitem{Berrehrah:2014kba} 
  H.~Berrehrah, P.~B.~Gossiaux, J.~Aichelin, W.~Cassing and E.~Bratkovskaya,
  Phys.\ Rev.\ C {\bf 90}, no. 6, 064906 (2014)
  doi:10.1103/PhysRevC.90.064906
  [arXiv:1405.3243 [hep-ph]].
  
  
\bibitem{Plumari:2011mk} 
  S.~Plumari, W.~M.~Alberico, V.~Greco and C.~Ratti,
  Phys.\ Rev.\ D {\bf 84}, 094004 (2011)
  doi:10.1103/PhysRevD.84.094004
  [arXiv:1103.5611 [hep-ph]].
  

\bibitem{vanHees:2007me} 
  H.~van Hees, M.~Mannarelli, V.~Greco and R.~Rapp,
  Phys.\ Rev.\ Lett.\  {\bf 100}, 192301 (2008)
  doi:10.1103/PhysRevLett.100.192301
  [arXiv:0709.2884 [hep-ph]].
  
\bibitem{Riek:2010fk} 
  F.~Riek and R.~Rapp,
  Phys.\ Rev.\ C {\bf 82}, 035201 (2010)
  doi:10.1103/PhysRevC.82.035201
  [arXiv:1005.0769 [hep-ph]].
  
\bibitem{Abreu:2011ic} 
  L.~M.~Abreu, D.~Cabrera, F.~J.~Llanes-Estrada and J.~M.~Torres-Rincon,
  Annals Phys.\  {\bf 326}, 2737 (2011)
  doi:10.1016/j.aop.2011.06.006
  [arXiv:1104.3815 [hep-ph]].

\bibitem{Berrehrah:2014tva} 
  H.~Berrehrah, P.~B.~Gossiaux, J.~Aichelin, W.~Cassing, J.~M.~Torres-Rfincon and E.~Bratkovskaya,
  Phys.\ Rev.\ C {\bf 90}, 051901 (2014)
  doi:10.1103/PhysRevC.90.051901
  [arXiv:1406.5322 [hep-ph]].
  
\bibitem{Ruggieri:2013bda} 
  M.~Ruggieri, F.~Scardina, S.~Plumari and V.~Greco,
  Phys.\ Lett.\ B {\bf 727}, 177 (2013)
  doi:10.1016/j.physletb.2013.10.014
  [arXiv:1303.3178 [nucl-th]].

 
\bibitem{Ruggieri:2013ova} 
  M.~Ruggieri, F.~Scardina, S.~Plumari and V.~Greco,
  Phys.\ Rev.\ C {\bf 89}, no. 5, 054914 (2014)
  doi:10.1103/PhysRevC.89.054914
  [arXiv:1312.6060 [nucl-th]].


\bibitem{Ferini:2008he} 
  G.~Ferini, M.~Colonna, M.~Di Toro and V.~Greco,
  Phys.\ Lett.\ B {\bf 670}, 325 (2009)
  doi:10.1016/j.physletb.2008.10.062
  [arXiv:0805.4814 [nucl-th]].

\bibitem{Greco:2008fs} 
  V.~Greco, M.~Colonna, M.~Di Toro and G.~Ferini,
  Prog.\ Part.\ Nucl.\ Phys.\  {\bf 62}, 562 (2009)
  doi:10.1016/j.ppnp.2008.12.029
  [arXiv:0811.3170 [hep-ph]].
    
\bibitem{Greco:2003xt} 
  V.~Greco, C.~M.~Ko and P.~Levai,
  Phys.\ Rev.\ Lett.\  {\bf 90}, 202302 (2003)
  doi:10.1103/PhysRevLett.90.202302
  [nucl-th/0301093].

\bibitem{Greco:2003mm} 
  V.~Greco, C.~M.~Ko and P.~Levai,
  Phys.\ Rev.\ C {\bf 68}, 034904 (2003)
  doi:10.1103/PhysRevC.68.034904
  [nucl-th/0305024].


\bibitem{Cacciari:2005rk} 
  M.~Cacciari, P.~Nason and R.~Vogt,
  Phys.\ Rev.\ Lett.\  {\bf 95}, 122001 (2005)
  doi:10.1103/PhysRevLett.95.122001
  [hep-ph/0502203].


\bibitem{Cacciari:2012ny} 
  M.~Cacciari, S.~Frixione, N.~Houdeau, M.~L.~Mangano, P.~Nason and G.~Ridolfi,
  JHEP {\bf 1210}, 137 (2012)
  doi:10.1007/JHEP10(2012)137
  [arXiv:1205.6344 [hep-ph]].
  
\bibitem{Bazavov:2011nk} 
  A.~Bazavov {\it et al.},
  Phys.\ Rev.\ D {\bf 85}, 054503 (2012)
  doi:10.1103/PhysRevD.85.054503
  [arXiv:1111.1710 [hep-lat]].
  
 \bibitem{Pet} C. Peterson {\it et al.},  Phys. Rev. D {\bf 27}, 105 (1983).
 
 \bibitem{hffs} D. Besson, {Eur. Phys. J. C}, {\bf 15} (2000), p. 218
 
 
\bibitem{GarciaRecio:2008dp} 
  C.~Garcia-Recio, V.~K.~Magas, T.~Mizutani, J.~Nieves, A.~Ramos, L.~L.~Salcedo and L.~Tolos,
  { Phys.\ Rev.\ D} {\bf 79} (2009) 054004
  
\bibitem{Gamermann:2010zz} 
  D.~Gamermann, C.~Garcia-Recio, J.~Nieves, L.~L.~Salcedo and L.~Tolos,
  {Phys.\ Rev.\ D} {\bf 81} (2010) 094016

\bibitem{Romanets:2012hm} 
  O.~Romanets, L.~Tolos, C.~Garcia-Recio, J.~Nieves, L.~L.~Salcedo and R.~G.~E.~Timmermans,
  { Phys.\ Rev.\ D} {\bf 85} (2012) 114032
  
\bibitem{GarciaRecio:2012db}
  C.~Garcia-Recio, J.~Nieves, O.~Romanets, L.~L.~Salcedo and L.~Tolos,
  { Phys.\ Rev.\ D} {\bf 87} (2013) 034032

\bibitem{Garcia-Recio:2013gaa}
  C.~Garcia-Recio, J.~Nieves, O.~Romanets, L.~L.~Salcedo and L.~Tolos,
  { Phys.\ Rev.\ D} {\bf 87} (2013) 074034

\bibitem{Tolos:2013gta} 
  L.~Tolos,
  Int.\ J.\ Mod.\ Phys.\ E {\bf 22}, 1330027 (2013)

\bibitem{Das:2013lra} 
  S.~K.~Das, S.~Ghosh, S.~Sarkar and J.~e.~Alam,
  Phys.\ Rev.\ D {\bf 88}, no. 1, 017501 (2013)
  doi:10.1103/PhysRevD.88.017501
  [arXiv:1303.2476 [nucl-th]].
  
\bibitem{Das:2015hla} 
  S.~K.~Das, F.~Scardina, S.~Plumari and V.~Greco,
  J.\ Phys.\ Conf.\ Ser.\  {\bf 668}, no. 1, 012051 (2016)
  doi:10.1088/1742-6596/668/1/012051
  [arXiv:1509.06307 [nucl-th]].
  
\bibitem{Tieulent:2015wec} 
  R.~Tieulent [ALICE Collaboration],
  arXiv:1512.02253 [nucl-ex].
  
\bibitem{Eskola:2008ca} 
  K.~J.~Eskola, H.~Paukkunen and C.~A.~Salgado,
  JHEP {\bf 0807}, 102 (2008)
  doi:10.1088/1126-6708/2008/07/102
  [arXiv:0802.0139 [hep-ph]].
  
  
\bibitem{Minissale:2015zwa} 
  V.~Minissale, F.~Scardina and V.~Greco,
  Phys.\ Rev.\ C {\bf 92}, no. 5, 054904 (2015)
  doi:10.1103/PhysRevC.92.054904
  [arXiv:1502.06213 [nucl-th]].
   
  
\end{thebibliography}
\end{document}